\documentclass[a4paper,12pt,reqno]{amsart}
\usepackage[utf8]{inputenc}
\usepackage[T1]{fontenc}
\usepackage{amsmath,amsxtra,amssymb,amsthm,amsfonts}

\usepackage{xfrac}
\usepackage{mathrsfs}
\usepackage{mathtools}   % loads »amsmath«
\usepackage{tikz}
\usepackage{xcolor}
\usetikzlibrary{arrows, automata}
\usepackage{color,hyperref}
\definecolor{dblue}{rgb}{0, 0, 0.72}
\numberwithin{equation}{section}
\newtheorem{lemma}{Lemma}[section]
\newtheorem{prop}[lemma]{Proposition}
\newtheorem{theorem}[lemma]{Theorem}

\newtheorem{cor}[lemma]{Corollary}

\newtheorem{prob}[lemma]{Problem}
\newtheorem{rem}[lemma]{Remark}
\newtheorem{remark}[lemma]{Remark}

\newtheorem{corollary}[lemma]{Corollary}

\newcommand{\re}{\begin{rem}\rm}
  \newcommand{\mar}{\end{rem}}
\newtheorem{exam}[lemma]{Example}

\newtheorem{defi}[lemma]{Definition}

\newcommand{\ee }{\mathrm{I}\!\!1}

\newcommand{\ketbra}[1]{|{#1}\rangle\langle{#1}|}

\renewcommand{\for}{\begin{eqnarray*}}

\newcommand{\mel}{\end{eqnarray*}}

\newcommand{\bE}{{\mathbb E}}

\newcommand{\bJ}{{\mathbb J}}

\newcommand{\bM}{{\mathbb M}}

\newcommand{\bR}{{\mathbb R}}

\newcommand{\cB}{{\mathcal B}}

\newcommand{\cD}{{\mathcal D}}
\newcommand{\cE}{{\mathcal E}}

\newcommand{\cH}{{\mathcal H}}

\newcommand{\cK}{{\mathcal K}}
\newcommand{\cL}{\mathcal{L}}

\newcommand{\cR}{{\mathcal R}}

\newcommand{\cU}{\mathcal{U}}

\DeclareMathOperator{\Tr}{Tr}
\DeclareMathOperator{\tr}{tr}

\DeclareMathOperator{\BKM}{BKM}

\DeclareMathOperator{\SLD}{SLD}
\DeclareMathOperator{\RLD}{RLD}

\newcommand{\pl}{\hspace{.1cm}}

\newcommand{\qd}{\end{proof}\vspace{0.5ex}}

\newcommand{\al}{\alpha}

\newcommand{\si}{\sigma}

\newcommand{\la}{\lambda}

\newcommand{\id}{\iota_{\infty,2}^n}

\newcommand{\pf}{\begin{proof}}

\newcommand{\be}{\left|{\atop}}

\newcommand{\xspace}{\hbox{\kern-2.5pt}}
\newcommand{\xyspace}{\hbox{\kern-1.1pt}}

\newcommand\bra[1]{\langle  #1|}
\newcommand\ket[1]{| #1\rangle}

\newcommand{\norm}[2]{\parallel \! #1 \! \parallel_{#2}}

\allowdisplaybreaks

\definecolor{LightGray}{rgb}{0.94,0.94,0.94}
\definecolor{VeryLightBlue}{rgb}{0.9,0.9,1}
\definecolor{LightBlue}{rgb}{0.8,0.8,1}
\definecolor{DarkBlue}{rgb}{0,0,0.6}
\definecolor{LightGreen}{rgb}{0.88,1,0.88}
\definecolor{MidGreen}{rgb}{0.6,1,0.6}
\definecolor{DarkGreen}{rgb}{0,0.6,0}
\definecolor{DarkGrreen}{rgb}{0,0.8,0}

\definecolor{VeryLightYellow}{rgb}{1,1,0.9}
\definecolor{LightYellow}{rgb}{1,1,0.6}
\definecolor{MidYellow}{rgb}{1,1,0.5}
\definecolor{DarkYellow}{rgb}{0.8,1,0.3}
\definecolor{VeryLightRed}{rgb}{1,0.9,0.9}
\definecolor{LightRed}{rgb}{1,0.8,0.8}
\definecolor{DarkRed}{rgb}{0.8,0.2,0}
\definecolor{DarkRedb}{rgb}{0.6,0.2,0}
\definecolor{DarkLila}{rgb}{0.8,0,1}
\definecolor{Beige}{rgb}{0.96,0.96,0.86}
\definecolor{Gold}{rgb}{1.,0.84,0.}
\definecolor{Goldb}{rgb}{0.7,0.3,0.5}
\definecolor{MyYellow}{rgb}{1.,0.84,0.8}

\newcommand{\lan}{\langle}
\newcommand{\ran}{\rangle}

\def\11{\mathbb{I}}

\usepackage{graphicx}

\DeclareRobustCommand\openone{\leavevmode\hbox{\small1\normalsize\kern-.33em1}}

\renewcommand{\id}{\rm{id}}
\renewcommand{\be}{\begin{equation}}
	\renewcommand{\ee}{\end{equation}}
\newcommand{\bea}{\begin{eqnarray}}
	\newcommand{\eea}{\end{eqnarray}}
\newcommand{\beas}{\begin{eqnarray*}}
	\newcommand{\eeas}{\end{eqnarray*}}

\newtheorem*{theorem*}{Theorem}
\newtheorem*{remark*}{Remark}
\newtheorem*{lemma*}{Lemma}
\newtheorem*{cor*}{Corollary}
\newtheorem*{note*}{Note}
\newtheorem*{prop*}{Proposition}
\newtheorem*{example*}{Example}

\newcommand{\WYD}{\text{WYD}}
\usepackage{tikz-cd}

\usepackage[margin=1in]{geometry}

\begin{document}

\title[Sufficient statistic and recoverability via QFI metrics]{Sufficient statistic and recoverability via Quantum Fisher Information metrics }

\author[L. Gao]{Li Gao}
\address{Department of Mathematics\\ University of Houston, Houston, TX 77204, USA
} \email[Li Gao]{lgao12@uh.edu}
\author[H. Li]{Haojian  Li}
\address{Zentrum Mathematik, Technische Universit\"{a}t M\"{u}nchen, 85748 Garching, Germany}
\email[Haojian  Li]{haojian.li@tum.de}
\author[I. Marvian]{Iman Marvian}
\email[Iman Marvian]{iman.marvian@duke.edu}
\author[C. Rouz\'e]{Cambyse Rouz\'e}
 \address{Zentrum Mathematik, Technische Universit\"{a}t M\"{u}nchen, 85748 Garching, Germany}
\email[Cambyse Rouz\'{e} ]{cambyse.rouze@tum.de}

\begin{abstract}
We prove that for a large class of quantum Fisher information, a quantum channel is sufficient for a family of quantum states, i.e., the input states can be recovered from the output,   if and only if the quantum Fisher information is preserved under the quantum channel. This class, for instance, includes Winger-Yanase-Dyson skew information. On the other hand, interestingly, the SLD quantum Fisher information, as the most popular example of  quantum analog of Fisher information, does not satisfy this property. 

Our recoverability result is obtained by studying Riemannian monotone metrics on the quantum state space, i.e. metrics monotone decreasing under the action of quantum channels, a property often called data processing inequality. For two quantum states, the monotone metric gives the corresponding quantum $\chi^2$ divergence. We obtain an approximate recovery result in the sense that, if the quantum $\chi^2$ divergence is approximately preserved by a quantum channel, then two states can be approximately recovered by the Petz recovery map.  We also obtain a universal recovery bound for the $\chi_{\frac{1}{2}}$ divergence. 

Finally, we discuss applications in the context of quantum thermodynamics and the resource theory of asymmetry.
\end{abstract}
\maketitle
%On the other hand, we observe such property does not holds for the monotone metric or quantum $\chi^2$ divergence corresponding to the SLD and RLD quantum Fisher information.

\section{Introduction}
Quantum metrology studies high-resolution measurements of physical parameters of quantum systems. In both classical and quantum metrology, the Fisher information plays an important role as a metric measuring the amount of information a system carries about a parameter $\theta$. The concept of Fisher information goes back to mathematical statistics \cite{fisher1922mathematical}:
let $(\Omega,\mu)$ be a probability space and $X(\theta):\Omega \to \bR$ be a family of random variables depending on an unknown parameter $\theta$. The Fisher information of $X$ at $\theta$ is defined as
\begin{align}\label{eq:cfisher}I_X(\theta):=\mathbb{E}\Big[\left.\big(\partial_\theta \log p_X(\theta)\big)^2 \right\vert\theta\Big]=\int_{\Omega} \frac{|\partial_\theta  p_X(\theta,\omega)|^2}{p_X(\theta,\omega)}\,d\mu(\omega),\end{align}
where $\omega\mapsto p_X(\theta,\omega )$ is the probability density function of $X(\theta)$ with respect to $\mu$. By the famous Cram\'er–Rao bound \cite{rao1945information,cramer2016mathematical}, the Fisher information gives a fundamental limit on the precision of parameter estimation: for any unbiased estimator $\hat{\theta}$ of $\theta$ \footnotemark \footnotetext{Here an unbiased estimator satisfies $\mathbb{E}(\hat{\theta}|\theta)=\theta$.}, it holds that
$\text{Var}(\hat{\theta})\ge 1/{I_X(\theta)}$. The Cram\'er–Rao bound has been extended to the quantum setting  \cite{holevo2011probabilistic, braunstein1994statistical,braunstein1996generalized}:
 for a family of quantum states
$\rho_\theta$ the variance of any unbiased estimator $\hat{\theta}$ of $\theta$  satisfies
\begin{align}\label{eq:Cramer-Rao} \text{Var}(\hat{\theta})\ge \frac{1}{I_{\operatorname{SLD},\rho}(\theta)}\pl.\end{align}
where
\begin{align}
I_{\operatorname{SLD},\rho}(\theta)= \tr(\dot{\rho}_\theta\, \bJ_{\rho_\theta}(\dot{\rho}_\theta))\pl,\label{eq:SLD}
\end{align}
and $\bJ_{\rho}$ is the inverse of the symmetric multiplication map $\bJ^{-1}_\rho(A)=\frac{1}{2}(\rho A+A\rho)$.
In the physics literature $I_{\operatorname{SLD}}$ is often called  the quantum Fisher information (QFI). Following the quantum statistics literature, we call this quantity
the symmetric logarithmic derivative (SLD) QFI.

 A nice property of the
quantum  Cram\'er–Rao bound with SLD QFI is that, similar to its classical version,  it is asymptotically achievable; that is,  given state $\rho(\theta)^{\otimes n}$ there exists a measurement for which the above bound becomes
tight in the limit $n\rightarrow\infty$ \cite{barndorff2000fisher, braunstein1994statistical}. This essentially follows from the fact that (i) both quantum and classical Fisher information are additive, and (ii) there exists a measurement on a single copy of $\rho(\theta)$, for which the classical
Fisher information of the outcome is equal to $I_{\operatorname{SLD},\rho}(\theta)$. Then, combining these facts  with the asymptotic achievability  of the classical Cram\'er–Rao bound, one can establish the achievability of the Quantum Cram\'er–Rao bound \footnote{As it was noted in \cite{barndorff2000fisher}, the measurement achieving this bound, in general, depends on the unknown parameter $\theta$. \cite{barndorff2000fisher} shows how this measurement can be determined by consuming a sublinear number of copies.} Hence
SLD QFI has a distinguished role in statistics that puts it on par with classical Fisher information.

\subsection{Sufficient Statistic}

Another important property of classical Fisher information is in the context of sufficient statistic.
For the family of random variables $X$ with distribution $P_X(\theta)$, a statistic $t=T(X)$
is called sufficient  for parameter $\theta$ if it contains all the information in $X$ about $\theta$, such that given $t=T(X)$, the random variable $X$ becomes independent of $\theta$.

It is known (see c.f. \cite{schervish2012theory}) that under certain regularity conditions, e.g., if $P_X(\theta)$ has full support for all $\theta$, then
\begin{align}\label{eq:sufficiency} t = T(X) \text{ is a sufficient statistic for $\theta$ } \Longleftrightarrow I_{T(X)}(\theta)=I_X(\theta)\,.\end{align}
%Note that $I_T(\theta)\le I_X(\theta)$ for any statistic $T=t(X)$ by the DPI of the Fisher information. Thus a statistic is sufficient if and only if the DPI of the Fisher information is saturated.

In the quantum setting, the notion of sufficient statistic can be defined in terms of recoverability with quantum channels:
 given a family of quantum states $\rho_\theta$,
a quantum channel $\Phi$ is sufficient for the family $\rho_\theta$ if and only if there exists a quantum channel $\cR$ such that $\cR\circ \Phi(\rho_\theta)=\rho_\theta$ for all $\theta$. Such $\cR$ is called a recovery map, which means the original family $\rho_\theta$ can be fully recovered from the channel output $\Phi(\rho_\theta)$. Jen{\v{c}}ov{\'a} and Petz \cite{jenvcova2006sufficiency} showed that such quantum sufficiency can be characterized via the relative entropy\footnotemark \footnotetext{For two quantum states with density operators $\rho$ and $\sigma$, $D(\rho\|\sigma)=\tr(\rho\log \rho-\rho\log \sigma)$ if $\operatorname{supp}(\rho)\subseteq \operatorname{supp}(\sigma)$, and is infinite otherwise.}: $\Phi$ is sufficient if and only if there exists some state $\sigma$ such that \begin{align}D(\rho_\theta\|\sigma)=D(\Phi(\rho_\theta)\|\Phi(\sigma))\pl,\pl \forall \theta \label{eq:REsufficiency}.\end{align} In this situation, Petz's work \cite{petz1988sufficiency,petz1986sufficient} showed that there is a canonical recovery map, \begin{align} \label{eq:petzmap} \cR_{\sigma,\Phi}(\cdot)=\sigma^{\frac12}\Phi^\dagger( \Phi(\sigma)^{{-}\frac12}\cdot \Phi(\sigma)^{{-}\frac12})\sigma^{\frac12},\end{align} called Petz recovery map.

%Despite the progress on (approximate) recoverability via entropic quantities, the statistic sufficiency of QFI is less explored.

\subsection{Failure of SLD QFI in characterizing sufficiency.}
A natural question is whether,  similar to the classical case, the sufficiency of statistics can be determined based on SLD QFI. In other words, does $I_{\operatorname{SLD},{\rho}}(\theta)=I_{\operatorname{SLD},\Phi({\rho})}(\theta)$ imply that there exists a recovery channel  $\cR$ such that
$\cR\circ \Phi(\rho_\theta)=\rho_\theta$?
Surprisingly, it turns out that the answer is negative.

\begin{prop}\label{Thm0}
There exists a smooth family of full-rank qubit state $\rho_\theta$ and a quantum channel $\Phi$,
\[ I_{\operatorname{SLD},{\rho}}(\theta)= I_{\operatorname{SLD},{\Phi(\rho)}}(\theta)\pl ,\pl  \forall \theta\]
yet there does not exist a recovery channel $\mathcal{R}$, such that  $\mathcal{R}\circ\Phi(\rho_\theta)=\rho_\theta$ for all $\theta$.
\end{prop}
Without the full-rank assumption, such non-recovery examples has been observed in \cite{marvian2022operational} (see \cite{kagan2005sufficiency,pollard2013note} for the counter-example to the full support assumption in the classical case). It was shown in \cite{marvian2022operational} that for any system $A$ with density operator $\rho_A$ and Hamiltonian $H_A$, there exists a purification  $|\psi\rangle_{AB}$ and Hamiltonian $H_B$ on the purifying system $B$,  such that the SLD QFI for the family of pure states  \[{|\psi(t)\rangle_{AB}=(e^{-i H_A t}\otimes e^{-i H_B t})   |\psi\rangle_{AB}: t\in\mathbb{R}}\] is equal to the SLD QFI  for the family of reduced density $\rho_A(t)=e^{-i H_A t} \rho_A  e^{i H_A t} $, which can be obtained from the first family by discarding system $B$.
However, despite preservation of SLD QFI under partial trace, it is impossible to recover the original state $|\psi(t)\rangle_{AB}$ from $\rho_A(t)$. Theorem \ref{Thm0} shows that this phenomenon also happens for smooth families of full-rank states, in contrast to the classical Fisher information \eqref{eq:sufficiency}.

\subsection{Sufficiency via regular QFI metrics} The failure of  SLD QFI in characterizing  sufficient statistic  motivates us to consider other quantum analogs of classical Fisher information that may satisfy this property. Indeed, quantum extensions of Fisher information in the context of information geometry   have been intensively studied in \cite{petz1996monotone,petz1996riemannian,lesniewski1999monotone,petz2011introduction,hayashi2002two,holevo2011probabilistic,kosaki2005matrix,petz2002covariance}.  \color{black} Viewing the space of all positive quantum states as a manifold, \eqref{eq:SLD} induces a Riemmanian metric, defined for any quantum state $\rho$ as
\begin{align}\gamma_\rho(A):= \tr(A\,\bJ_{\rho}(A)) \pl, \end{align}
for all traceless, Hermitian operators $A$, interpreted as tangent vectors at $\rho$. This metric is of special interests, as it is monotone under any quantum channel $\Phi$,
\begin{align} \label{eq:monotonicity}\gamma_\rho(A)\ge  \gamma_{\Phi(\rho)}(\Phi(A))\pl. \end{align}
Such an inequality is commonly called data processing inequality (DPI). %Indeed, the data processing inequality of the standard relative entropy and more general quantum $f$-divergences can follow from the monotonicity of corresponding metrics \cite{lesniewski1999monotone}.

In the classical setting, it was proved by \v{C}encov \cite{vcencov1978algebraic} that the Fisher information metric is the unique Riemannian metric (up to scaling) satisfying the monotonicity \eqref{eq:monotonicity}. From this perspective, there are more than one quantum analog of the classical Fisher information whose corresponding metrics satisfy DPI. This family of metrics, called monotone metrics or quantum Fisher information (QFI) metrics, were first proposed by \v{C}encov and Morozova \cite{morozova1989markov}, and later fully classified by Petz \cite{petz1996monotone}. It was observed by Lesniewski and Ruskai \cite{lesniewski1999monotone} that any QFI metric correspond to the Hessian of a given quantum $f$-divergence. %, so that the DPI \eqref{eq:monotonicity} is inherited from DPI for quantum $f$-divergences.
In the special case where all the density operators in the family are diagonal in a fixed basis, QFI metrics all reduce to the classical Fisher information of the probability distribution defined by the eigenvalues of the density operators.

  Interestingly, it turns out that SLD QFI in \eqref{eq:SLD} is indeed the smallest QFI, which explains its special role in the  Cram\'er–Rao bound. There is also a largest QFI, namely the Right-Logarithmic Derivative (RLD) Fisher information defined by \begin{align}I_{\operatorname{RLD},\rho}(\theta)=\tr(\rho^{-1}(\theta)|\dot{\rho}(\theta)|^2) \pl.\label{eq:RLD}\end{align}

While SLD QFI fails to characterize sufficient statistics, our first main result shows that a large class of QFI metrics, which we call them ``regular'' metrics,   characterize sufficiency (see Section \ref{sec:monotone} for the definition of regular QFI metrics).

\begin{theorem}\label{thm:b}
Given a smooth family of quantum states $(\rho_\theta)_{\theta\in (a,b)}$ with full support, a quantum channel $\Phi$ is sufficient for $\theta$ if and only if
\[ I_{\rho}(\theta)=I_{\Phi(\rho)}(\theta)\pl, \quad \pl \forall \theta\in (a,b)\ ,\]
holds for all/any {\bf regular} quantum Fisher information $I$. In particular, for any $o\in(a,b)$ in this family, the corresponding Petz recovery map of state $\rho_o$ defined in Eq.(\ref{eq:petzmap}) recovers\footnote{As we show in Theorem \ref{Thm5}, this result also holds for rotated Petz maps.}  the full family, i.e.,
\[ \mathcal{R}_{\rho_o,\Phi}\circ\Phi(\rho_\theta)=\rho_\theta, \quad \pl \forall \theta\in (a,b)\ .\] \end{theorem}
The following well-known metrics are all regular QFI and therefore, according to our theorem, characterize sufficient statistic.

  \begin{itemize}

\item[a)]  Wigner-Yanase-Dyson (WYD) skew information: Given a density operator $\rho$ and Hermitian operator $H$, the WYD skew information is defined as
\begin{align} W_{H}^{(\al)}(\rho)=-\frac{1}{2}\tr([\rho^\al, H][\rho^{1-\al},H])\pl ,\pl   0<\al<1 \pl. \label{eq:WYD}\end{align}
$W_{\al,H}(\rho)$ is the QFI of the family $\rho_t=e^{-i tH} \rho e^{itH},t\in \mathbb{R}$ as the Petz-R\'enyi divergence $Q_\al(\rho,\sigma)= \tr(\rho^{1-\al}\sigma^\al)$.

\item [b)] $x^\al$-metric: for $0< \al< 1$,
\[ I_{\al,\rho}(\theta)=\tr(\dot{\rho}_\theta\rho^{-\al}\dot{\rho}_\theta \rho^{\al-1}) \pl, \]
for the special case $\al=\frac{1}{2}$, we call it symmetric inverse QFI.

\item [c)] Bogoliubov-Kubo-Mori (BKM) Fisher information: BKM QFI is the negative Hessian of the relative entropy, defined as
\begin{align}\label{eq:BKM}
I_{\BKM,\rho}(\theta)&:=-\left.\frac{\partial^2}{\partial \theta_1 \partial \theta_2 }D(\rho_{\theta_1}\|\rho_{\theta_2})\right|_{\theta_1=\theta_2=\theta}\\ &=\int_{0}^\infty\tr( \dot{\rho}_\theta (\rho_\theta+r1)^{-1}\dot{\rho}_\theta(\rho_\theta+r1)^{-1})\,dr\pl,
\end{align}
is a regular QFI satisfying the theorem above.

  \end{itemize}

In Section \ref{Sec:Applications} we discuss implications of this result in the context of quantum thermodynamics and the resource theory of asymmetry. In conclusion, while to this date most applications of QFI in physics have been based on the special case of SLD QFI,
our results clearly demonstrate operational and physical relevance of general QFI metrics, beyond this special case.

\subsection{Approximate Recoverability}

%Approximate recoverability

%Petz's works has stimulated a lot of research on sufficiency of quantum channels \cite{wang2022revisiting,jenvcova2006sufficiency,jenvcova2006sufficiency2,ohya2004quantum}.
Over the last decade,  a series of works established a stronger notion of recoverability, namely approximate recoverability \cite{fawzi2015quantum,junge2018universal,sutter2016universal,sutter2017multivariate,carlen2020recovery,gao2021recoverability}.
This line of research was initiated by a work of  Fawzi and Renner  \cite{fawzi2015quantum} on approximate quantum Markov chains. The notion of approximate recoverability has  found various applications in different areas of physics, including high energy physics and condensed matter theory \cite{cotler2019entanglement,hayden2021markov,hayden2019learning}.  A notable result is by Junge \emph{et al} \cite{junge2018universal}, who proved that
\begin{align}\label{eq:approximate1}
D(\rho\|\si)- D(\Phi(\rho)\|\Phi(\si))\ge -2\log F(\rho,\mathcal{R}_{\sigma,\Phi}^{\operatorname{uni}}\circ \Phi(\rho))\ge \norm{\rho-\mathcal{R}_{\sigma,\Phi}^{\operatorname{uni}}\circ \Phi(\rho)}{1}^2,\vspace{-5pt}
\end{align}
where $F(\rho,\sigma):=\|\sqrt{\rho}\sqrt{\sigma}\|_1$ is the fidelity between two quantum states $\rho,\sigma$ and $\norm{\cdot}{1}$ denotes the trace norm. Here $\mathcal{R}_{\sigma,\Phi}^{\operatorname{uni}}$ is called the universal recovery map given by the integral
\begin{align}\label{eq:universal} \cR_{\sigma,\Phi}^{\operatorname{uni}}=\int_{\mathbb{R}}\cR_{\sigma,\Phi}^{\frac{t}{2}} \,d\beta(t)\pl,\qquad  \pl d\beta(t)=\frac{\pi}{2(\cosh (\pi t)+1)}\,dt\end{align}
where $\cR^t_{\sigma,\Phi}$ is the rotated version of Petz Recovery map,
\begin{align}\label{eq:rotated}\cR^t_{\sigma,\Phi}(\cdot)=\sigma^{-it}\cR_{\sigma,\Phi}\big( \Phi(\sigma)^{it}\cdot \Phi(\sigma)^{-it}\big)\sigma^{it}\pl.\end{align}

Despite the progress on approximate recoverability via entropic quantities, it remains open whether it can be characterized using QFI. Our second main result address this question.

\begin{theorem}\label{thm:c2}Suppose $\rho_\theta\ge \lambda 1$ for some $\lambda>0$ and all $\theta\in (a,b)$, then for any $s<r$
\[ \lambda^{-\frac{1}{2}}\int_{s}^r\sqrt{I_{\operatorname{BKM},\rho}(\theta)-I_{\operatorname{BKM},\Phi(\rho)}(\theta)}\,d\theta\ge  D(\rho_r\|\rho_s)-D(\Phi(\rho_r)\|\Phi(\rho_s)) \pl.\]
\end{theorem}
Combining this with the existing
results on approximate recoverability in terms of relative entropy, such as
Eq.(\ref{eq:approximate1}), one can obtain recovery bounds in terms of BKM metric.

%Second, in the case of $\frac{1}{2}$-metric we establish the following bound
%\begin{align}
%\gamma_\rho^s(A)-\gamma_{\Phi(\rho)}^s(\Phi(A))\ge\gamma_{\rho}^s(A-\mathcal{R}_{\rho,\Phi}\circ\Phi(A))\ge  \norm{A-\mathcal{R}_{\rho,\Phi}\circ\Phi(A)}{1}^2\ ,
  %  \end{align}
%where $\rho$ is an arbitrary  density operator $\rho\in \cD_+(\cH)$, $A\in \mathcal{B}(\mathcal{H})$  [IM: is $\rho$ full rank?].
%Note that $A-\mathcal{R}_{\rho,\Phi}\circ\Phi(A)$ is the  derivative of the error term $\rho(\theta)-\mathcal{R}_{\rho,\Phi}\circ\Phi(\rho(\theta))$ with respect to $\theta$, and therefore this result establishes a bound on the norm of this derivative.

\subsection{Quantum $\chi^2$ divergence} The proof of the above (approximate) recoverability results follows from a simpler setting, namely that of quantum $\chi^2$ divergence.   The classical $\chi^2$ divergence of two distributions $P$ and $Q$ is defined by
\begin{align}\label{eq:chi2c}\chi^2(P,Q)=\bE_Q\left|\frac{dP}{dQ}-1\right|^2\ .\end{align}
For two quantum states $\rho,\sigma$ and a given QFI metric $\gamma$, the quantum analog of $\chi^2$ divergence is
\begin{align}\label{eq:chi2}\chi^2(\rho,\sigma)= \gamma_\sigma(\rho-\sigma)\pl,\end{align}
which can be understood as the QFI for the linear interpolation family $\rho_t=t \sigma+(1-t)\rho$.  In the special case where $\rho$ and $\sigma$ commute this quantity reduces to the classical $\chi^2$ divergence in Eq.(\ref{eq:chi2c}) for the distributions defined by the eigenvalues of $\rho$ and $\sigma$.
The above definition based on QFI metrics guarantees that the $\chi^2$ divergence associated to each QFI metric inherits the nice properties of the metric, such as monotonicity under DPI.  These $\chi^2$ divergences have found applications in characterizing the mixing time of a quantum Markov process \cite{temme2010chi,gao2022complete2}.

It is natural to ask whether the recoverability or the stronger notion of approximate recoverability can be characterized with quantum $\chi^2$ divergences or the corresponding monotone metric $\gamma$, as Hessians of quantum $f$-divergence. Again, the answer can be negative or positive depending on
the choice of the metric $\gamma$. In the case of regular quantum $\chi^2$ divergences, i.e., those corresponding to regular QFI we show that
\begin{theorem}\label{thm:c}
Given two states $\rho$ and $\sigma$ with $\operatorname{supp}(\rho)\subseteq \operatorname{supp}(\sigma)$, a quantum channel $\Phi$ is sufficient for $\{\rho,\sigma\}$ if and only if
\[\chi^2(\rho,\sigma)=\chi^2(\Phi(\rho),\Phi(\sigma))\pl,\]
holds for any/all {\bf regular} quantum $\chi^2$-divergence. Moreover, if $\sigma>\lambda I$, there exists an explicit constant $K(\chi^2,\lambda)$ depending on $\chi^2$ and $\lambda$   such that for all $t\in \mathbb{R}$,
\[ \chi^2(\rho,\sigma)-\chi^2(\Phi(\rho),\Phi(\sigma))\ge \frac{\pi}{\cosh(\pi t)}K(\chi^2,\lambda)\norm{\rho-\cR_{\sigma,\Phi}^t\Phi(\rho)}{1}^2,\]
where $\cR^t_{\sigma,\Phi}$
 is the rotated Petz map defined in \eqref{eq:rotated}.
\end{theorem}
On the other hand, the above property does not generally hold for  non-regular
$\chi^2$-divergences, even though they still satisfy DPI inequality. Two notable examples are the SLD and RLD $\chi^2$-divergences ($\bJ_\sigma$ is as in \eqref{eq:SLD})
\[\chi_{\SLD}^2(\rho,\sigma)=\tr(\bJ_\sigma(\rho)^2\sigma)-1 \pl, \pl \chi_{\RLD}^2(\rho,\sigma)=\tr(\rho^2\sigma^{-1})-1\pl.\]

In the special case of symmetric inverse metric \[ \chi_{\frac12}^2(\rho,\sigma)=\tr((\rho-\sigma)\sigma^{-1/2}(\rho-\sigma)\sigma^{-1/2})=\tr(\rho\sigma^{-\frac{1}{2}}\rho\sigma^{-\frac{1}{2}})-1,\]
which is related to the Sandwiched 2-R\'enyi relative entropy ${D_2(\rho\|\sigma)=\log \tr(\rho\sigma^{-\frac{1}{2}}\rho\sigma^{-\frac{1}{2}})}$. The $\chi_{\frac12}^2$ divergence is known to enjoy some special properties (see \cite{cao2019tensorization}). Here, we achieve a universal approximate recovery bound.
\begin{theorem}\label{thm:d}
For two states $\rho$ and $\sigma$,
\[\chi_{\frac12}^2(\rho,\sigma)-\chi_{\frac12}^2(\Phi(\rho),\Phi(\sigma))\ge \norm{\rho-\cR_{\sigma,\Phi}\circ \Phi(\rho)}{1}^2\pl,\]
where $\cR_{\sigma,\Phi}$ is the Petz recovery map defined in \eqref{eq:petzmap}.
\end{theorem}
The above bound improves the result of \cite{cree2022approximate} by removing a state dependent constant $\norm{\sigma^{-1}}{}$, which can be unbounded in infinite dimensions. It also applies to classical Fisher information as all quantum $\chi^2$-divergence reduces to the classical $\chi^2$-divergence in the commutative setting. We also note that here our recovery map $\cR_{\sigma,\Phi}$ is the original Petz map, while it remains open whether the recovery bound \eqref{eq:approximate1} for relative entropy can be achieved with $\cR_{\sigma,\Phi}$.

\subsection{Applications: Quantum Thermodynamics and the resource theory of asymmetry}\label{Sec:Applications}
An important application of QFI is in the context of quantum thermodynamics and the closely-related   resource theory of asymmetry. QFI metrics provide a useful way of quantifying the amount of coherence of a system with respect to its energy eigenbasis and, more generally,  the amount of asymmetry (symmetry-breaking) of the system with respect to a given symmetry group \cite{Marvian_thesis, marvian2014extending,  girolami2014observable, yadin2016general, kwon2018clock,  marvian2022operational}.
Below we illustrate the application of our results for coherence, and refer to Section 6 for the more general setting of asymmetry.

For a system with density operator $\rho$ and Hamiltonian $H$, consider the family of time evolution of this system, namely states $\rho(t)=e^{-i H t}\rho e^{i H t}$ for $t\in\mathbb{R}$. For any QFI metric $\gamma$, the QFI of this family with respect to the time parameter $t$ is time-independent, that is, for any $t\in\mathbb{R}$
\begin{equation}\label{IM:def1}
I_{\rho}(t)=\gamma_{\rho(t)}(\dot{\rho}(t), \dot{\rho}(t))=\gamma_\rho(i[H, \rho], i [H, \rho]):=I_{H}(\rho)\pl.
\end{equation} This quantity determines the asymmetry  of the system with respect to the time translation symmetry, or equivalently, the  energetic coherence of the system with respect to the eigenbasis of Hamiltonian $H$. In particular, $I_{H}(\rho)=0$ if and only if $[\rho,H]=0$, namely, the state is diagonal in the energy-eigenbasis. %Furthermore, it is additive for tensor product states of non-interacting systems, where the total Hamiltonian is the sum of the Hamiltonians of the individual systems.

An important and useful property of this function is its monotonicity under any quantum operation $\mathcal{E}$ (CPTP map) that respects the covariance condition
\begin{equation}\label{cov1}
  \mathcal{E}\big(e^{-i H_\text{in} t} (\cdot) e^{i H_\text{in} t}  \big)=e^{-i H_\text{out} t}\mathcal{E}\big( \cdot\big) e^{i H_\text{out} t}  \ , \pl \forall \pl  t\in\mathbb{R},
\end{equation}
where $H_\text{in}$ and $H_\text{out}$ are Hamiltonians for the input and output system, respectively. Operations satisfying this property are sometimes called time-translation invariant operations.
This property implies that under channel $\mathcal{E}$ the family of states $e^{-i H_\text{in} t}\rho e^{i H_\text{in} t}$ is mapped to  $e^{-i H_\text{out} t}\mathcal{E}(\rho) e^{i H_\text{out} t}$.
Then,  the DPI
for QFI metrics immediately implies that $I_H$  is monotone under any such map $\mathcal{E}$, \[I_{H_{\text{out}}}(\mathcal{E}(\rho))\le I_{H_{\text{in}}}(\rho)\pl.\]  Any function satisfying this monotonicity is called  a measure of asymmetry with respect to the symmetry group under consideration, which in this case is the time translation symmetry.\footnote{It is often also required that a measure of asymmetry should vanish for all states that are invariant under the action of symmetry, which in this case are states satisfying $e^{-i H t}\rho e^{iH t}=\rho$ for all $t\in\mathbb{R}$. It can be easily seen that function $I_H$ satisfies this property.}

While all QFI metrics can be used to quantify asymmetry and coherence, recent work \cite{marvian2022operational} has singled out SLD QFI, as the measure of asymmetry with an operational interpretation: namely, it quantifies the \emph{coherence cost} of preparing a general mixed state from pure coherent states.  RLD QFI has also shown to be useful for characterizing the distillation of energetic coherence, i.e., time-translation asymmetry \cite{marvian2020coherence}.

%Previous works have studied measures of asymmetry such as SLD and RLD QFI. In particular,

The present work reveals that, in addition to SLD and RLD QFI,
regular QFI metrics are particularly useful for quantifying asymmetry and energetic coherence. In particular, they can determine whether the resourcefulness   of the system has been degraded by noise or any process that respects the symmetry. We show that %More precisely, suppose under a CPTP map $\mathcal{E}$ satisfying Eq.(\ref{cov1}), a system with state $\rho$ and Hamiltonian $H$ is evolved  to another system with state $\rho'$ and Hamiltonian $H'$. Is this process reversible via symmetric operations? Our result address this question, at least in the case of full-rank density operators:

\begin{theorem}\label{IM:Thm1}
Consider systems $A$ and $B$ with Hamiltonians $H_A$ and $H_B$, respectively. Let $\rho\in \cB(\cH_A)$ be a full-rank density operator and let $\cE:\cB(\cH_A)\to \cB(\cH_B)$ be time-translation invariant quantum channel. Then, there exists a time-translation invariant channel $\mathcal{R}$ such that $\mathcal{R}(\cE(\rho))=\rho$, if and only if
\begin{align} I_{H_A}(\rho)=I_{H_B}(\cE(\rho))\pl, \label{IM:cons1}\end{align}
where  $I_H$  is the QFI defined in Eq.(\ref{IM:def1}) with respect to some/all regular QFI metric $\gamma$.
\end{theorem}

An important example of regular QFI
metric is Wigner-Araki-Yanase skew information
\begin{align}
W^{(\alpha)}_H(\rho):=\Tr(\rho H^2)- \Tr(\rho^{1-\alpha} H \rho^{\alpha} H )\ ,
 \end{align}
for $0  < \alpha<1$, which have been previously studied as a measure of coherence and asymmetry \cite{Marvian_thesis, marvian2014extending, takagi2019skew}. It is worth mentioning that  the above property of regular QFI metrics, that their conservation implies reversibility, was  shown \cite{marvian2016clocks}  to hold for the relative entropy of asymmetry (also known as the asymmetry).\footnote{Unlike metrics is not additive in tensor-product states,  the relative entropy of asymmetry grow logarithmically with the number of copies.}

Theorem \ref{IM:Thm1} follows from applying Theorem \ref{thm:b} for the family $\rho(t)=e^{-i H_A t}\rho e^{i H_A t}$ with  $t\in\mathbb{R}$. By the time-translation invariance of $\cE$, $\cE(\rho(t))=e^{-i H_B t}\cE(\rho) e^{i H_B t}: t\in\mathbb{R}$. Furthermore, since QFI $I_\rho(t)=I_{H_A}(\rho)$ is time-independent, Eq.(\ref{IM:cons1}) implies that the QFI preserves under $I_\rho(t)=I_{H_A}(\rho)=I_{H_B}(\rho)= I_{\cE(\rho)}(t)$ for all $t\in \mathbb{R}$. Then, our Theorem \ref{thm:b} implies that the Petz recovery of $\rho$ recovers the states for all time $t$, i.e.,
\begin{align}\label{IM:rec3}
\mathcal{R}_{\rho,\mathcal{E}}(\cE(\rho(t)))=\rho(t)\ ,\  \forall t\in\mathbb{R}
\end{align}
Furthermore, for any finite $T>0$  the time-averaged version of Petz map
\begin{align}\label{eq:Ravg}
\mathcal{R}_{\text{avg}, T}(\cdot)=\frac{1}{T}\int_{-T/2}^{T/2}dt\  e^{i H_A t}\  \mathcal{R}_{\rho,\mathcal{E}}\left(e^{-i H_B t} (\cdot)e^{i H_B t}\right)\ e^{-i H_A t}\ ,
\end{align}
also satisfies \eqref{IM:rec3} (Note that $\rho(t+s)= e^{i H_A t}\rho(s) e^{-i H_A t}$ and similar for $\cE(\rho(t))$). In general, $\mathcal{R}_{\text{avg}, T}$ is not covariant. Nevertheless, since for finite-dimensional Hilbert space $\cH_A$ and $\cH_B$, the set of CPTP maps is compact, there exists a limit point
$\mathcal{R}_{\text{avg}, \infty}=\lim_{n\rightarrow \infty} \mathcal{R}_{\text{avg}, T_n}$. It is straightforward to show that  $\mathcal{R}_{\text{avg}, \infty}$  satisfies the time translation invariant condition Eq.(\ref{cov1}). This proves the Theorem \ref{IM:Thm1}. We refer to Section 6 for the more general cases of asymmetry of compact Lie groups. \\

\noindent {\bf Outline of the rest of the paper.} We first discuss the non-recoverability of SLD and RLD QFI metric in the Section 2. We briefly review in Section 3 the definitions of general quantum monotone metrics, Fisher information and $\chi^2$ divergences.  Section 4 is devoted to the recoverability of regular $\chi^2$ divergences (Theorem \ref{thm:c}) and the universal recoverability bound for $\chi_{1/2}^2$ (Theorem \ref{thm:c}). Based on that we prove the recoverability (Theorem \ref{thm:b}) and approximate recoverability (Theorem \ref{thm:c2}) of regular QFI in Section 5. Section 6 discusses the application of our results in quantum coherence and asymmetry.\\

\noindent {\bf Acknowledgement.}
LG is partially supported by NSF grant DMS-2154903. IM is supported by
NSF grants FET-1910571, Phy-2046195, FET-2106448.  C.R. acknowledges
the support of the Munich Center for Quantum Sciences and Technology, funding from the Humboldt Foundation as well as by the Deutsche Forschungsgemeinschaft (DFG,
German Research Foundation) under Germanys Excellence Strategy EXC-2111 390814868. H.L. acknowledges support by the DFG cluster of excellence 2111 (Munich
Center for Quantum Science and Technology).\\

\noindent{\bf Notations.} We write $\bM_n$ for the set of $n\times n$ complex matrices.
Given a finite dimensional Hilbert space $\mathcal{H}$,
% and $\mathcal{H}$ be finite dimensional Hilbert spaces.
we denote $\mathcal{B}(\mathcal{H})$, $\cB(\cH)_{\operatorname{sa}}$ and $\cB(\cH)_+$ as the set of bounded, Hermitian and positive (semi-definite) operators respectively. We write $\langle A,B \rangle=\tr(A^{*}B)$ for the Hilbert-Schmidt inner product, where $\tr$ stands for the standard matrix trace. The Schatten norm of order $p\ge 1$ is defined as $\|A\|_p:=\tr(|A|^p)^{1/p}$ and $\mathcal{S}_p(\cH)$ denotes the Schatten-$p$ space.
We denote by $\mathcal{D}(\mathcal{H})$ the subset of density operators (positive semi-definite and trace $1$) on $\mathcal{H}$, $\mathcal{D}_+(\mathcal{H})$ by the subset of invertible density operators on $\mathcal{H}$. We use $I$ for the identity operator in $\cB(\cH)$ and $\id$ for the identity map on $\cB(\cH)$. We write $A^*$ as the adjoint of an operator $A$ and $\Phi^\dagger$ as the adjoint of a map $\Phi$ with respect to Hilbert-Schmidt inner product. Given two finite dimensional Hilbert spaces $\cH$ and $\cK$, a quantum channel $\Phi:\cB(\cH)\to \cB(\cK)$ is a completely positive trace preserving map. In particular, $\Phi({\cD(\cH)})\subset {\cD(\cK)}$ preserves the density operators.\\

%\color{red}
%Recycle? in recent years QFI metrics have also found  applications in various areas of quantum physics, such as quantum metrology \cite{demkowicz2020multi,jiang2014quantum,frowis2016detecting,rath2021quantum,liu2015quantum}, uncertainty principle \cite{luo2000quantum, gibilisco2007uncertainty}

%entanglement detection \cite{li2013entanglement}.

% The proof of our Theorem \ref{thm:b} is based on the following recovery results for quantum $\chi^2$ divergences.

%Petz's works has stimulated a lot of research on sufficiency of quantum channels \cite{wang2022revisiting,jenvcova2006sufficiency,jenvcova2006sufficiency2,ohya2004quantum}.

%recent progress \cite{faulkner2022approximate,junge2020multivariate}
%\color{black}

%\newpage

\section{Insuffcient statistic preserving SLD quantum Fisher information}\label{Sec:counterexample}
 In this section, we present the counter-example that quantum sufficiency cannot be characterized via SLD QFI.
To show this it is useful to compare SLD
and RLD QFI. For a family of states $\rho_\theta$, let $\mathbb{J}^{-1}_{\rho_\theta}(\dot{\rho}_{\theta})=L_{\theta}$. Recall that
%\[ I_{\operatorname{RLD},\rho}(\theta)=\tr(\rho_\theta^{-1}|\dot{\rho}_\theta|^2)\pl, \pl I_{\operatorname{SLD},\rho}(\theta)=\tr(L_\theta^2 \rho_\theta)\pl,\] where $L(\theta)$ is the symmetric logarithmic derivative, i.e., the Hermitian operator satisfying \begin{align} \dot{\rho}_\theta=\frac{1}{2}(L_\theta\rho_\theta+\rho_\theta L_\theta)\ . \end{align}
\[  I_{\operatorname{RLD},\rho}(\theta)=\tr(\rho_\theta^{-1}|\dot{\rho}_\theta|^2)\pl, \pl I_{\operatorname{SLD},\rho}(\theta)=\tr(\dot{\rho}_\theta\mathbb{J}_{\rho_{\theta}}(\dot{\rho}_\theta)=\tr(L_{\theta}^{2}\rho_{\theta})\pl.
\]

Note that $I_{\operatorname{RLD}}$ is infinite whenever $\text{supp}(|\dot{\rho}_\theta|)\subseteq \text{supp}(\rho_\theta)$, but $I_{\operatorname{SLD}}$ can be finite as long as $ \text{supp}(\dot{\rho}_\theta)^\perp\subseteq \text{supp}(\rho_\theta)^\perp$.
Moreover, RLD QFI can be rewritten as
\begin{align}
I_{\operatorname{RLD},\rho}(\theta)&=\tr(\rho^{-1}_\theta|\dot{\rho}_\theta|^2)=\frac{1}{4} \tr\Big(\rho^{-1} (L\rho+\rho L)^2\Big)\\ &= \frac{3}{4} I_{\operatorname{SLD},\rho}(\theta)+ \frac{1}{4}\tr(\rho^{-1} L\rho^2 L)   \ ,
\end{align}
where to simplify the notation we have omitted the parameter $\theta$ in the second equation. We conclude that
\begin{align}\label{gap}
I_{\operatorname{RLD},\rho}(\theta)-I_{\operatorname{SLD},\rho}(\theta)=\frac{1}{4}P_{L_\theta}(\rho_\theta)   \ ,
\end{align}
where the quantity
\begin{align}
P_L(\rho)=\tr(\rho^{-1} L\rho^2 L)-\Tr(\rho L^2)=-\tr(\rho^{-1} [\rho,L]^2)\ ,
\end{align}
 is indeed the RLD QFI for the family of states $e^{it L} \rho e^{-it L}$ with respect to parameter $t$, called the \emph{purity of coherence} of $\rho$ with respect to $L$ \cite{marvian2020coherence}. In particular, $P_L(\rho)\ge 0$ and is zero if and only if $[\rho,L]=0$ commute.

%In summary, at point $\theta$ the gap between RLD and SLD QFI vanishes  if, and only if, $\rho(\theta)$ commutes with the symmetric logarithmic derivative $L(\theta)$. [IM: is this stronger than $[\rho, \rho]$]
Fix a parameter value $\theta_o$ and let $\mathcal{L}_o$ be the pinching map that dephases its input with respect to the spectrum of $\rho_{\theta_o}$.
 Define $\sigma_\theta=\mathcal{L}_o(\rho_\theta)$. Note that
\[ \dot{\sigma}_\theta=\mathcal{L}_o(\dot{\rho}_\theta)=\frac{1}{2}\mathcal{L}_o(L_{\theta_o}\rho_{\theta_o}+\rho_{\theta_o}L_{\theta_o})=\frac{1}{2}(L_{\theta_o}\sigma_{\theta_o}+\sigma_{\theta_o}L_{\theta_o}) \]
Thus, $L_o$ remain the SLD for $\sigma_\theta$ at $\theta=\theta_o$ and the SLD QFI at $\theta_o$ does not change under this map, i.e.,
\begin{align}
I_{\operatorname{SLD},\sigma}(\theta_o)=\tr(L^2_{{\theta_o}} \sigma_{\theta_o})= \tr(L^2_{\theta_o} \rho_{\theta_o})= I_{\operatorname{SLD},\rho}(\theta_o)  \ .
\end{align}
Furthermore, the state  $\sigma_{\theta_o}$ commutes with $L_{\theta_o}$ and therefore $P_{L_{\theta_o}}(\sigma_{\theta_o})=0$, which means the gap between RLD and SLD QFI vanishes, i.e.,
\begin{align}
I_{\operatorname{RLD},\sigma}(\theta_o)=I_{\operatorname{SLD},\sigma}(\theta_o)= I_{\operatorname{SLD},\rho}(\theta_o)  \ .
\end{align}
Note that  unless $[L_{\theta_o}, \rho_{\theta_o}]= 0$, then $I_{\operatorname{SLD},\rho}(\theta_o)<I_{\operatorname{RLD},\rho}(\theta_o)$ and hence $\cL_o$ is not sufficient because $I_{\operatorname{RLD},\sigma}(\theta_o)<I_{\operatorname{RLD},\rho}(\theta_o)$.
In summary, we observe
\begin{prop}
For a family of states $\rho_\theta$, the difference between RLD and SLD QFI equal to $P_{L_\theta}(\rho_\theta)$, which is zero if and only if $[L_\theta, \rho_\theta]=0$, where $L_\theta$ is the symmetric logarithm derivative of $\rho_\theta$.

Fix a parameter value value $\theta_o$, the dephasing map $\cL_o$ relative to the eigen-subspaces of $L_{\theta_o}$ always preserved SLD QFI at $\theta_o$, but strictly decrease RLD QFI if $[L_{\theta_o},\rho_{\theta_o}]\neq 0$, hence not recoverable.
\end{prop}
In general, because SLD operator $L_{\theta_o}$ depends on the parameter $o$, so does the map $\mathcal{L}_o$. However, as we will see in the following example, it is possible
to have a family of states $\rho_\theta$ such that the SLD operator $L_{\theta_o}$ share the same eigen-subspaces, which means the dephasing map $\mathcal{L}_\theta=\mathcal{L}$ is independent of $\theta$. Furthermore, this family can be chosen to be full-rank. Such a family satisfies the full-rank condition in Theorem \ref{thm:b} while SLD QFI remains conserved under the map $\mathcal{L}$, but its conservation does not imply sufficiency of the output statistic. We emphasis that this contrasts with the classical case,
where conservation of Fisher information for probability distributions with full-support implies sufficiency \cite[Theorem 2.8]{schervish2012theory}.

\subsection*{Qubit example:}  Consider the family of qubit density operators
\[
\rho_\theta=\left(
\begin{array}{cc}
p(\theta)  &   \epsilon \times r(\theta)  \\
\epsilon \times r(\theta) &  1-p(\theta)
\end{array}
\right): \ \
\theta\in [a, b]\ ,
\]
defined for $\theta\in[a, b]$,
where  $p: [a, b]\rightarrow (0,1)$  is an arbitrary function with finite, non-zero derivative  $\dot{p}(\theta)$. The function $r$ is determined by $p$ via equation
\begin{align}
r(\theta)=\exp \int^{\theta}_{a}\  \frac{\dot{p}(s)(1-2 p(s))}{2p(s)(1-p(s))} ds\ ,
\end{align}
and $\epsilon$ is chosen such that
\begin{align}
0< \epsilon^2 < \min_{\theta\in [a,b]} \frac{p(\theta)(1-p(\theta)) }{r^2(\theta) }\ .
\end{align}
The latter condition guarantees that $\rho_\theta$ is positive and full-rank for all $\theta\in[a,b]$.

Then, one can easily check that
the Hermitian operator
 \[
L_\theta=\left(
\begin{array}{cc}
\dot{p}(\theta)/p(\theta)  &  0  \\
0 &  -\dot{p}(\theta)/({1-p(\theta)})
\end{array}
\right) \ ,
\]
satisfies the equation
\begin{equation}
\dot{\rho}_\theta=\frac{1}{2}\Big[\rho_\theta L_\theta+L_\theta \rho_\theta\Big]\ ,
\end{equation}
and therefore is the SLD of the family $\rho_\theta$. Note that $L_\theta$ is always diagonal. Furthermore,
the assumption that $\dot{p}(\theta)\neq 0$ implies that $L(\theta)$ in non-degenerate.
Let  $\mathcal{L}$ be the dephasing map in $\{|0\rangle,|1\rangle \}$ basis. Applying this map to state $\rho_\theta$  we obtain the family of states
 \[
\sigma_\theta=\mathcal{L}(\rho_\theta)=\left(
\begin{array}{cc}
p(\theta)  &   0 \\
0 &  1-p(\theta)
\end{array}
\right): \ \
\theta\in [a, b]\ .
\]
Under this dephasing, SLD QFI remains conserved. However, because the original family $\rho_\theta$ is not diagonal in  $\{|0\rangle,|1\rangle \}$ basis, RLD QFI for the family of states $\sigma_\theta$ is strictly less than the RLD QFI for the family of states
 $\rho_\theta$. Therefore, even though SLD QFI is preserved under $\cL$, the RLD QFI decreases strictly. Hence $\sigma_\theta=\cL(\rho_\theta)$ is not a sufficient statistic for the original family $\rho_\theta$.

\section{Monotone metrics}\label{sec:monotone}
The faithful state space $\cD:=\cD_+(\cH)$ is a submanifold in the real Euclidean space $\cB(\cH)$. At each point $\rho\in \cD$, the tangent space
\[T_\rho\cD=\{ A\in \cB(\cH) \pl |\pl  A=A^*\pl, \pl \tr(A)=0\}\]
is the subspace of traceless Hermitian operators. A Riemannian metric on $\cD$ is a smooth assignment $\rho\mapsto\gamma_\rho$ to a positive bilinear form $\gamma_\rho:T_\rho\cD\times T_\rho\cD\to \mathbb{R}$.
\begin{defi}\label{def:gamma}We say a Riemannian metric $\gamma$ is a monotone metric, if for any quantum channel $\Phi:\cB(\cH)\to \cB(\cK)$,
\[\gamma_\rho (A,A)\ge \gamma_{\Phi(\rho)} (\Phi(A),\Phi(A))\pl, \pl\quad  \forall A\in T_\rho\cD\pl.\]
We will use the short notation $ \gamma_\rho^g(A):=\gamma_\rho^g(A,A)$.
\end{defi}
The monotone metrics are classified by  operator anti-monotone (i.e. decreasing) functions $g:(0,\infty)\to (0,\infty)$.
Given  $\rho\in\mathcal{D}$, we define 
\[ \bJ_\rho^g=g(L_\rho R_\rho^{-1})R_\rho^{-1}\pl, \]
where $L_\rho(A)=\rho A, R_\rho(A)=A\rho$ are left and right multiplications respectively.
Based on the work of Chentsov and Morozova \cite{morozova1989markov},Petz in \cite{petz1996monotone} proved that every monotone metric admits the following form
\begin{align}\label{eq:gmetric} \gamma_\rho^g(A,B)=\lan A, \bJ_\rho^g (B)\ran= \lan A \rho^{-\frac{1}{2}},g(L_\rho R_\rho^{-1}) (B\rho^{-\frac{1}{2}})\ran\pl.
\end{align}
If  $\rho=\sum_{i}\lambda_i\ketbra{\phi_i}$, the metric is explicit with matrix coefficients
\[ \gamma_\rho (A,A)=\sum_{i,j}c(\lambda_i,\lambda_j)|\bra{\phi_i} A\ket{\phi_j}|^2\]
where $c(x,y)=y^{-1}g(xy^{-1})$ is called Morozova-Chentsov function. If we assume that the definition of $\gamma_\rho^g$ for $A,B\in \cB(\cH)$ has the symmetric property
\[\gamma_\rho(A,B)=\gamma_\rho(B^*,A^*),\]
this corresponds to operator anti-monotone functions satisfying
\begin{align} g(x^{-1})=xg(x)\pl, \pl g(1)=1\pl. \label{eq:symmetric}\end{align}
These $\bJ_\rho^g$ operators are inverses to operator means between $L_\rho$ and $R_\rho$, and admit the following integral form (see \cite{lesniewski1999monotone})
\begin{align}\label{eq:integral} \bJ_\rho^g=R_{\rho}^{-1}\int_{0}^\infty\frac{1}{s+\Delta_\rho} \,\nu_g(s)\,ds\end{align}
where $\Delta_\rho=L_\rho R_\rho^{-1}$ is the relative modular operator and $\nu_g(s)\,ds$ is a finite positive measure satisfying $\nu_{g}(s^{-1})=s\nu_{g}(s)$.

 We will encounter the following special cases in our discussion.
\begin{exam}{\rm (1) SLD metric: for $g(x)=\frac{2}{x+1}$ and $\nu_g=2\delta_1$ being the point mass at $1$,
\begin{align} \label{eq:Bures} \gamma_\rho^{\text{SLD}}(A,B)=2\tr(A^*(L_\rho+R_\rho)^{-1}B)\pl.
\end{align}
$\gamma^{ \operatorname{SLD}}$ is also called Bures metric in the literature. \\
(2) RLD metric: for $\nu_g=\delta_0$ being point mass at $s=0$
\begin{align} \label{eq:Right} \gamma_\rho^{\operatorname{RLD}}(A,B)=\frac{1}{2}\tr(A^*B\rho^{-1})
\end{align}
This corresponds to the RLD Fisher information introduced in \eqref{eq:RLD}.\\
(3) BKM metric: for $g(x)=\frac{\log x}{x-1}$ and $\nu_g(s)=\frac{1}{s+1}$,
\begin{align} \label{eq:BKMmetric} \gamma_\rho^{\BKM}(A,B)=\int_{0}^\infty\tr(A^*(\rho+s1)^{-1}B(\rho+s1)^{-1})\,ds\,.
\end{align}
(4) $x^\al$-metrics: for $g(x)=\frac{1}{2}(x^{-\al}+x^{\al-1}),\al\in (0,1)$ and $\nu_g(s)=\frac{\sin\pi s}{2\pi}\big(s^{-\al}+s^{\al-1}  \big)$,
\begin{align} \label{eq:almetric} \gamma_\rho^{(\al)}(A,B)=\tr(A^*\rho^{-\al} B\rho^{\al-1})\ .
\end{align}
A special case is $\al=\frac{1}{2}$, which we write as 
\begin{align} \label{eq:12metric} \gamma_\rho^{s}(A,B):=\tr(A^*\rho^{-\frac{1}{2}} B\rho^{-\frac{1}{2}} )\,.
\end{align}
(5) WYD metric: for $\al\in (0,1)$, $g(x)=\frac{(1-x^\al)(1-x^{1-\al})}{\al(1-\al)(1-x^2)}$ and $\nu_g(s)=\frac{\sin(\pi s)}{\pi}(1+s)^{(\al-2)}+\frac{\sin(\pi s^{-1})}{s\pi}(1+s^{-1})^{(\al-2)}$,
\begin{align} \label{eq:wydmetric} \gamma_\rho^{\WYD}(A,B)=  \frac{\partial^2}{\partial s\partial t}\tr\big((\rho+sA)^\al(\rho+tB)^{1-\al} \big)|_{s=t=0}
\end{align}
This gives the WYD skew information as in Eq. \eqref{eq:WYD}. 
}
\end{exam}

It is clear from functional calculus that if $g_1\le g_2$, then
\[\gamma_{\rho}^{g_1}(A,A)\le \gamma_{\rho}^{g_2}(A,A)\pl, \pl \forall A\in \cB(\cH)\pl.\]
From this perspective, for any monotone metric $\gamma^g$,
\[ \gamma_\rho^{\text{Bures}}(A,A)\le \gamma_\rho^{g}(A,A)\le \gamma_\rho^{\text{R}}(A,A)\]
and the $\frac12$-metric is the smallest among all $\al$-metrics.
\begin{defi}\label{def:QFI}We say a monotone metric $\gamma_g$ or its associated operator anti-monotone function $g$ is regular if the Lebesgue measure $ds$ is absolutely continuous with respect to $\nu_g(s)\,d(s)$.
\end{defi}
The metric $\gamma^{\BKM}$, $\gamma^{(\al)}$ and $\gamma^{\WYD}$ are regular but $\gamma_\rho^{\text{Bures}}$ and $\gamma_\rho^{\operatorname{RLD}}$ are not. We will see later that this is the reason for the non-recoverability for the latter two.

We note that for  a general density operator $\rho$, the monotone metric $\gamma^g$ are well-defined and finite for $A$ with $s(A)\le s(\rho)$, where $s(\rho)$ is the support of $\rho$. For example, the RLD metric $\gamma_\rho^{\RLD}(A)=+\infty$ as long as $s(A)\nleq s(\rho)$. If in addition, $\lim_{x\to 0^+}g(x)$ exists and finite, $\gamma^g$ is also finite for self-adjoint $A$ if $(1-s(\rho))A(1-s(\rho))=0$. This is the case for SLD metric.
\newpage

\section{Recovery via monotone metrics}
\subsection{Exact Recovery via monotone metrics}
Let $\gamma^g$ be a monotone metric. For two quantum states $\rho,\sigma\in \cD(\cH)$, the quantum $\chi^2$ divergence is
\begin{align} \chi^2_g(\rho,\sigma):=\gamma_\sigma^g(\rho-\sigma)\pl. \label{eq:chi2}\end{align}
It satisfies the data processing inequality: for any quantum channel $\Phi$,
\[\chi^2_g(\rho,\sigma)\ge \chi^2_g(\Phi(\rho),\Phi(\sigma))\,.\] The main goal of this section is to prove that for a regular $g$, \[\chi^2_g(\rho,\sigma)= \chi^2_g(\Phi(\rho),\Phi(\sigma)) \Longleftrightarrow \rho=\mathcal{R}_{\sigma,\Phi}^{t}\circ \Phi(\rho)\]
for any/all $t\in \mathbb{R}$, where $\mathcal{R}_{\sigma,\Phi}^{t}$ is the rotated Petz map of $\sigma$ and channel $\Phi$:
\begin{align}\label{eq:rotatedpetz} \mathcal{R}_{\sigma,\Phi}^{t}(A)=\sigma^{\frac12-it}\Phi^{\dagger}(\Phi(\sigma)^{-\frac12+it}A\Phi(\sigma)^{-\frac12-it})\sigma^{\frac12+it}.
\end{align}
Indeed, we will show a quantitative version of the above result. Our argument is inspired from \cite{CV20a} and \cite{gao2021recoverability}.

\medskip

We start with a lemma on the Stinespring dilation of the rotated Petz map.
\begin{lemma}\label{lemma:isoine}
Let $\rho\in\mathcal{D}_{+}(\mathcal{H})$ and $\Phi:\mathcal{B}(\mathcal{H})\to\mathcal{B}(\mathcal{K})$ be a quantum channel. For any $t\in\mathbb{R}$, we define the linear map
\begin{align*}
V_{\rho,t}: \mathcal{B}(\mathcal{K}) \to \mathcal{B}(\mathcal{H}),\qquad  V_{\rho,t}(A)=\Phi^{\dagger}(A\Phi(\rho)^{-\frac{1}{2}-it})\rho^{\frac{1}{2}+it},
\end{align*}
Then
\begin{enumerate}
\item[i)]$V^{*}_{\rho,t}V_{\rho,t}$ is a contraction on $\mathcal{S}_2(\cH)$, i.e. $V_{\rho,t}^{*}V_{\rho,t}\le I$.
\item[ii)] $V_{\rho,t}^{*}\Delta_{\rho}V_{\rho,t}\leq \Delta_{\Phi(\rho)}$ as positive operators on $\mathcal{S}_2(\cH)$
\end{enumerate}
\end{lemma}
\begin{proof}  For any $A\in\mathcal{B}(\mathcal{H})$, we have
\begin{align*}
\langle A, V_{\rho,t}^{*}V_{\rho,t}(A) \rangle&=\langle V_{\rho,t}(A), V_{\rho,t}(A) \rangle\\
&=\langle \Phi^{\dagger}(A\Phi(\rho)^{-\frac{1}{2}-it})\rho^{\frac{1}{2}+it},\Phi^{\dagger}(A\Phi(\rho)^{-\frac{1}{2}-it})\rho^{\frac{1}{2}+it} \rangle \\
&=\tr\left(\rho^{\frac{1}{2}-it}\Phi^{\dagger}(\Phi(\rho)^{-\frac{1}{2}+it}A^{*})\Phi^{\dagger}(A\Phi(\rho^{-\frac{1}{2}-it}))\rho^{\frac{1}{2}+it}\right)\\
&\leq \tr\left(\rho\Phi^{\dagger}\left(\Phi(\rho)^{-\frac{1}{2}+it}A^{*}A\Phi(\rho)^{-\frac{1}{2}-it}\right)\right)\\
&=\langle \Phi(\rho), \Phi(\rho)^{-\frac{1}{2}+it}A^{*}A\Phi(\rho)^{-\frac{1}{2}-it} \rangle\\
&=\langle A,A \rangle,
\end{align*}
where the above inequality follows from the operator Schwarz inequality
$$\Phi^{\dagger}(X^{*})\Phi^{\dagger}(X)\leq \Phi^{\dagger}(X^{*}X),\quad \forall X\in\mathcal{B}(\mathcal{H}).$$
Similarly,
\begin{align*}
\langle A, V_{\rho,t}^{*}\Delta_{\rho}V_{\rho,t}(A) \rangle=&\langle V_{\rho,t}(A), \Delta_{\rho}V_{\rho,t}(A)
\rangle\\
=&\langle \Phi^{\dagger}(A\Phi(\rho)^{-\frac{1}{2}-it})\rho^{\frac{1}{2}+it},\rho \Phi^{\dagger}(A\Phi(\rho)^{-\frac{1}{2}-it})\rho^{\frac{1}{2}+it}\rho^{-1}) \rangle\\
=&\tr\Big(\Phi^{\dagger}(\Phi(\rho)^{-\frac{1}{2}+it}A^{*})\rho\Phi^{\dagger}(A\Phi(\rho)^{-\frac{1}{2}-it})\Big)\\
=&\tr\Big(\rho \Phi^{\dagger}(A\Phi(\rho)^{-\frac{1}{2}-it})\Phi^{\dagger}(\Phi(\rho)^{-\frac{1}{2}+it}A^{*})\Big)\\
\leq &\tr\Big(\rho \Phi^{\dagger}(A\Phi(\rho)^{-1}A^*)\Big)\\
=&\langle A, \Delta_{\Phi(\rho)}(A) \rangle,
\end{align*}
where again the inequality follows from the operator Schwarz inequality.
\end{proof}

\noindent Our next lemma is a modification of \cite[Lemma 2.1]{CV20a}.

\begin{lemma}\label{lemma:contraction}
%Let $\Delta\in \cB(\cH)_+$ be a positive operator and $V:\cK\to \cH$ be a contraction. If $\Delta_0\ge V^*\Delta V$, then for any $s>0$ and $h\in \cH$,\\ \\
Let $\Delta:\cB(\cH)\to \cB(\cH)$, $\tilde{\Delta}:\cB(\cK)\to \cB(\cK), V: \cB(\cK)\to \cB(\cH)$ be positive linear maps and $V^{*}V$ be a contraction. If $\tilde{\Delta}\geq V^{*}\Delta V$ as positive operators on $\mathcal{S}_{2}(\cH)$, then for any $s\geq 0$ and $h\in \cB(\cH)$,
\begin{align*}
\langle h,\,(s+\Delta)^{-1}(h)\rangle-\langle h,\,V(s+\tilde{\Delta})^{-1}V^* (h) \rangle\geq \langle h_s,\,(s
+\Delta)(h_s)\rangle,
\end{align*}
where \begin{align}h_{s}=(s+\Delta)^{-1}h-V(s+\tilde{\Delta})^{-1}V^{*}(h)\pl. \label{eq:cv20ast} \end{align}
\end{lemma}
\begin{proof}
Let us first calculate the right hand side using the definition of $h_{s}$,
\begin{align*}
\langle h_{s},\,(s+\Delta)(h_s)\rangle&=\langle h,\,(s+\Delta)^{-1}(h) \rangle-2\langle h,\,V(s+\tilde{\Delta})^{-1} V^{*}(h) \rangle\\
&+\langle V(s+\tilde{\Delta})^{-1}V^{*} (h),\,(s+\Delta)\,V(s+\tilde{\Delta})^{-1}V^{*}(h)\rangle.
\end{align*}
Then since $V^{*}V\leq I$ and $V^{*}(\Delta+s)V\leq (\tilde{\Delta}+s)$ by the assumption, we obtain an upper bound of the last term
\begin{align*}
\langle
&V(s+\tilde{\Delta})^{-1}V^{*}(h),(s+\Delta)V(s+\tilde{\Delta})^{-1}V^{*}(h)\rangle\\
&\qquad\qquad= \langle (s+\tilde{\Delta})^{-1}V^{*}(h),\, V^{*}(s+\Delta)V(s+\tilde{\Delta})^{-1}V^{*}(h)\rangle \\
&\qquad\qquad\leq  \langle (s+\tilde{\Delta})^{-1}V^{*}(h),\,(s+\tilde{\Delta})(s+\tilde{\Delta})^{-1}V^{*}(h)\rangle\\
&\qquad\qquad= \langle h,\,V(s+\tilde{\Delta})^{-1}V^{*}(h) \rangle\pl,
\end{align*}
  which yields the desired inequality in the lemma.
\end{proof}

Applying the above lemmas to $V_{\rho,t}$ and $\Delta_\rho$, we have
\begin{align*}
\langle A,\,(s+\Delta_{\rho})^{-1}(A)\rangle-\langle A,\,V_{\rho,t}(s+\Delta_{\Phi(\rho)})^{-1}V_{\rho,t}^{*} (A) \rangle\geq \langle A_{s,t},\,(s
+\Delta_{\rho})(A_{s,t})\rangle,
\end{align*}
where \begin{align}A_{s,t}=(s+\Delta_{\rho})^{-1}A-V_{\rho,t}(s+\Delta_{\Phi(\rho)})^{-1}V_{\rho,t}^{*}(A). \label{eqcv20ast} \end{align}

The next lemma shows the above expression upper bounds the approximate recoverability.

\begin{lemma} \label{12recov}
 Let $\rho\in\mathcal{D}_{+}(\mathcal{H})$ and $\Phi:\cB(\cH)\to \cB(\cK)$ be a quantum channel. Let $\mathcal{R}_{\rho,\Phi}^{t}$ be as defined in \eqref{eq:rotatedpetz}.
	\begin{align*}
		\|A-\mathcal{R}_{\rho,\Phi}^{t}(\Phi(A))\|_{1}\le \left\|\Big(A-\mathcal{R}_{\rho,\Phi}^{t}(\Phi(A))\Big)\rho^{-1/2}\right\|_{2}\leq \frac{\cosh(\pi t)}{\pi}\, \left\|\int_{0}^{\infty}s^{-\frac{1}{2}+it}\Delta_{\rho}^{\frac{1}{2}}\tilde{A}_{s,t}\,ds\right\|_{2}\,,
	\end{align*}
	where \begin{align}\tilde{A}_{s,t}=(s+\Delta_{\rho})^{-1}(A\rho^{-\frac{1}{2}+it})-V_{\rho,t}(s+\Delta_{\Phi(\rho)})^{-1}V_{\rho,t}^{*}(A\rho^{-\frac{1}{2}+it}).\label{eqcv20ast1}
	\end{align}

\end{lemma}
\begin{proof} By the integral representation $\displaystyle r^{-\frac{1}{2}+it}=\frac{\cosh(\pi t)}{\pi}\int_{0}^{\infty} \frac{s^{-\frac{1}{2}+it}}{r+s}ds$ from \cite{Kom66}, we have
\begin{align*}
&\frac{\cosh(\pi t)}{\pi}\int_{0}^{\infty}s^{-\frac{1}{2}+it}\Delta_{\rho}^{\frac{1}{2}} \tilde{A}_{s,t}ds\\
&\qquad\qquad\qquad=\Delta_{\rho}^{\frac{1}{2}}\frac{\cosh(\pi t)}{\pi}\int_{0}^{\infty}s^{-\frac{1}{2}+it}(s+\Delta_{\rho})^{-1}(A\rho^{-\frac{1}{2}+it})ds\\
&\qquad\qquad\qquad\qquad-\Delta_{\rho}^{\frac{1}{2}}V_{\rho,t}\frac{\cosh(\pi t)}{\pi}\int_{0}^{\infty}s^{-\frac{1}{2}+it}(s+\Delta_{\Phi(\rho)})^{-1}V_{\rho,t}^{*}(A\rho^{-\frac{1}{2}+it})ds\\
&\qquad\qquad\qquad=\Delta_{\rho}^{it}(A\rho^{-\frac{1}{2}+it})-\Delta_{\rho}^{\frac{1}{2}}V_{\rho,t}\Delta_{\Phi(\rho)}^{-\frac{1}{2}+it}V_{\rho,t}^{*}(A\rho^{-\frac{1}{2}+it})\\
&\qquad\qquad\qquad=\rho^{it}A\rho^{-\frac{1}{2}}-\rho^{\frac{1}{2}}\Phi^{\dagger}(\Phi(\rho)^{-\frac{1}{2}+it}\Phi(A)\Phi(\rho)^{-\frac{1}{2}-it})\rho^{it}\\
&\qquad\qquad\qquad=\rho^{it}(A-\mathcal{R}_{\rho,\Phi}^{t}(\Phi(A)))\rho^{-\frac{1}{2}}.
\end{align*}
Then the inequality follows from the H\"older inequality,
\begin{align*}
&\|\rho^{it}(A-\mathcal{R}_{\rho,\Phi}^{t}(\Phi(A)))\rho^{-\frac{1}{2}}\|_{2}\|\rho^{\frac{1}{2}}\|_{2}\geq \|A-\mathcal{R}_{\rho,\Phi}^{t}(\Phi(A))\|_{1}\pl. \qedhere
\end{align*}
\end{proof}

We shall now estimate the above recoverability bounds via the monotone metrics. This is our main technical lemma.
\begin{lemma}\label{1abg}
Let $\gamma^g$ be a monotone metric given by the integral representation \eqref{eq:integral} with measure $\nu_g(s)ds$.
	For any $0<a\le b<\infty$, suppose there exists a function $w:[0,\infty)\to\mathbb{R}_+$ such that
	\begin{align*}
	    W_{a,b}:=\int_a^b \frac{w(s)}{s}\,ds\,<\infty
	\end{align*}
	and a constant $C_g(a,b)>0$ such that on the interval $[a,b]$
	\begin{align*}
		\frac{1}{w(s)}\,ds\leq C_g(a,b)\nu_{g}(s)ds.
	\end{align*}
	Then we have for any $\rho\in \cD_+(\cH), A\in \mathcal{B}(\mathcal{H})$ and $t\in\mathbb{R}$
	
	\begin{align*}
		\|A-\mathcal{R}_{\rho,\Phi}^{t}(\Phi(A))\|_{1}\leq \,\frac{\cosh(\pi t)}{\pi}\left( 4\sqrt{a}h_{1}+\frac{4}{\sqrt{b}}h_{2}+C_g(a,b)^{\frac{1}{2}}W_{a,b}^{\frac{1}{2}} h_{3}\right),
	\end{align*}
	where
	\begin{align*}
 &h_{1}:=\tr(A^{*}\rho^{-1}A)^{\frac{1}{2}},
	\,h_{2}:=\frac{1}{2}\Big(\tr(\rho^{-2}A^{*}\rho A)^{\frac{1}{2}}+\tr(\Phi(\rho)^{-2}\Phi(A)^{*}\Phi(\rho) \Phi(A))^{\frac{1}{2}}\Big) \, \\ & h_{3}:=\left(\gamma_{\rho}^{g}(A)-\gamma_{\Phi(\rho)}^{g}(\Phi(A))\right)^{\frac{1}{2}}\,.
	\end{align*}
 %\begin{align*}
 %h_{2}:=\frac{1}{2}\Big(\left(\tr(\rho^{-2}A^{*}\rho A)\right)^{\frac{1}{2}}+\left(\tr(\Phi(\rho)^{-2}\Phi(A)^{*}\Phi(\rho) \Phi(A))\right)^{\frac{1}{2}}\Big)
 %\end{align*}
%	For invertible and positive definite $\rho\in\mathcal{B}(\mathcal{H})$, then
%	\begin{align*}
%		\|A-\mathcal{R}_{\rho,\Phi}^{t}(A)\|_{1}\leq \tr(\rho)\frac{\cosh(\pi t)}{\pi}\left( 4\sqrt{a}h_{1}+\frac{4h_{2}}{\sqrt{b}}+C(a,b,g)^{\frac{1}{2}}(\ln(b)-\ln(a)+b-a)^{\frac{1}{2}} h_{3}\right).
%	\end{align*}
\end{lemma}
\begin{proof} By Lemma \ref{12recov}, it is enough to bound $\|\int_{0}^{\infty}s^{-\frac{1}{2}+it}\Delta_{\rho}^{\frac{1}{2}}\tilde{A}_{s,t}ds\|_{2}$ from above. We split the integral into three terms
	\begin{align*}
		\int_{0}^{\infty}s^{-\frac{1}{2}+it}\Delta_{\rho}^{\frac{1}{2}}\tilde{A}_{s,t}ds=&\int_{0}^{a}s^{-\frac{1}{2}+it}\Delta_{\rho}^{\frac{1}{2}}\tilde{A}_{s,t}ds+\int_{a}^{b}s^{-\frac{1}{2}+it}\Delta_{\rho}^{\frac{1}{2}}\tilde{A}_{s,t}ds+\int_{b}^{\infty}s^{-\frac{1}{2}+it}\Delta_{\rho}^{\frac{1}{2}}\tilde{A}_{s,t}ds\\
		=&\text{I}+\text{II}+\text{III}.
	\end{align*}
	Let $h_{t,a}(x)=\int_{0}^{a}\frac{s^{-\frac{1}{2}+it}x^{\frac{1}{2}}}{(x+s)}ds$.
	Then
	\begin{align*}
		\|\text{I}\|_{2}&\leq\|h_{t,a}(\Delta_{\rho})(A\rho^{-\frac{1}{2}+it})\|_{2}+\left\|\int_{0}^{a}s^{-\frac{1}{2}+it}\Delta_{\rho}^{\frac{1}{2}}V_{\rho,t}(s+\Delta_{\Phi(\rho)})^{-1}V_{\rho,t}^{*}(A\rho^{-\frac{1}{2}+it}) ds\right\|_{2}.
	\end{align*}
	Next, we notice that
	\begin{align*}
		&\left\|\int_{0}^{a}s^{-\frac{1}{2}+it}\Delta_{\rho}^{\frac{1}{2}}V_{\rho,t}(s+\Delta_{\Phi(\rho)})^{-1}V_{\rho,t}^{*}(A\rho^{-\frac{1}{2}+it}) ds\right\|_{2}^{2}\\
		&\qquad=\langle V_{\rho,t}^{*}\Delta_{\rho}V_{\rho,t} \int_{0}^{a}\frac{s^{-\frac{1}{2}+it}}{s+\Delta_{\Phi(\rho)}}V_{\rho,t}^{*}(A\rho^{-\frac{1}{2}+it}) ds,\int_{0}^{a}\frac{s^{-\frac{1}{2}+it}}{s+\Delta_{\Phi(\rho)}}V_{\rho,t}^{*}(A\rho^{-\frac{1}{2}+it}) ds \rangle\\
		& \qquad\overset{(1)}{\leq}\langle \Delta_{\Phi(\rho)}
		\int_{0}^{a}\frac{s^{-\frac{1}{2}+it}}{s+\Delta_{\Phi(\rho)}}V_{\rho,t}^{*}(A\rho^{-\frac{1}{2}+it}) ds,\int_{0}^{a}\frac{s^{-\frac{1}{2}+it}}{s+\Delta_{\Phi(\rho)}}V_{\rho,t}^{*}(A\rho^{-\frac{1}{2}+it}) ds
		\rangle \\
  &\qquad\leq \langle\int_{0}^{a}\frac{s^{-\frac{1}{2}+it}\Delta_{\Phi(\rho)}^{\frac{1}{2}}}{s+\Delta_{\Phi(\rho)}}V_{\rho,t}^{*}(A\rho^{-\frac{1}{2}+it}) ds,\int_{0}^{a}\frac{s^{-\frac{1}{2}+it}\Delta_{\Phi(\rho)}^{\frac{1}{2}}}{s+\Delta_{\Phi(\rho)}}V_{\rho,t}^{*}(A\rho^{-\frac{1}{2}+it}) ds     \rangle\\
		&\qquad=\|h_{t,a}(\Delta_{\Phi(\rho)})V_{\rho,t}^{*}(A\rho^{-\frac{1}{2}+it})\|_{2}^{2},
	\end{align*}
	where the inequality in $(1)$ above follows from Lemma \ref{lemma:isoine}.
	Moreover, for $x>0$, we have
	\begin{align*}
		|h_{t,a}(x)|\leq \int_{0}^{a}\left|\frac{s^{-\frac{1}{2}+it}x^{\frac{1}{2}}}{(x+s)}\right|ds=  2\arctan{\sqrt{\frac{a}{x}}}\leq 2\,\sqrt{\frac{a}{x}}.
	\end{align*}
	Together with the inequality above and another use of Lemma \ref{lemma:isoine}, we can hence bound $\|I\|_{2}$ from above:
	\begin{align*}
		\|\text{I}\|_{2}&\leq 2\sqrt{a} \langle A\rho^{-\frac{1}{2}+it},\,\Delta_{\rho}^{-1}(A\rho^{-\frac{1}{2}+it}) \rangle^{\frac{1}{2}} +2\sqrt{a}\langle V_{\rho,t}^*(A\rho^{-\frac{1}{2}+it}),\,\Delta_{\Phi(\rho)}^{-1}V_{\rho,t}^* (A\rho^{-\frac{1}{2}+it}) \rangle^{\frac{1}{2}} \\
		&=2\sqrt{a}\tr(A^{*}\rho^{-1}A)^{\frac{1}{2}}+2\sqrt{a}\tr(\Phi(A^{*})\Phi(\rho)^{-1}\Phi(A))^{\frac{1}{2}}\\
		&\leq 4\sqrt{a}\tr(A^{*}\rho^{-1}A)^{\frac{1}{2}},
	\end{align*}
	where the last inequality follows
from the Schwarz-type operator inequality (see \cite{LR74}, \cite[Theorem 5.3]{wolf2012quantum}):
	\begin{align*}
		\Phi(A^{*})\Phi(\rho)^{-1}\Phi(A)\leq \Phi(A^{*}\rho^{-1}A).
	\end{align*}
(or simply the monotonicity of Right operator mean metric).
	Let us now consider $\|\text{III}\|_2$. Define $\displaystyle \tilde{h}_{t,b}(x)=\int_{b}^{\infty} \frac{x^{\frac{1}{2}}s^{-\frac{1}{2}+it}}{(x+s)}ds$. Then for $x>0$,
\[|\tilde{h}_{t,b}(x)|\le\int_{b}^{\infty} \left|\frac{x^{\frac{1}{2}}s^{-\frac{1}{2}+it}}{(x+s)}\right|ds=2\arctan(\sqrt{\frac{x}{b}})\le 2\sqrt{\frac{x}{b}} \]
Similar to I, we obtain that
	\begin{align*}
		\|\text{III}\|_{2}\leq \frac{2}{\sqrt{b}}\tr(\rho^{-2}A^{*}\rho A)^{\frac{1}{2}}+\frac{2}{\sqrt{b}}\tr(\Phi(\rho)^{-2}\Phi(A^{*})\Phi(\rho)\Phi(A))^{\frac{1}{2}}\pl.
	\end{align*}
	Finally, we consider the second integral:
	\begin{align*}
		\|\text{II}\|_{2}&\leq\int_{a}^{b} \sqrt{\frac{1}{s}}\,\|\Delta_{\rho}^{\frac{1}{2}}\tilde{A}_{s,t}\|_{2}\,ds\\
		&\leq \left(\int_{a}^{b}\frac{w(s)}{s}ds\right)^{\frac{1}{2}}\left(\int_{a}^{b}
		\frac{1}{w(s)}\|\Delta_{\rho}^{\frac{1}{2}}\tilde{A}_{s,t}\|_{2}^{2}\,ds\right)^{\frac{1}{2}}\\
		&= W_{a,b}^{\frac{1}{2}}\,\left(\int_{a}^{b}
		\frac{1}{w(s)}\|\Delta_{\rho}^{\frac{1}{2}}\tilde{A}_{s,t}\|_{2}^{2}\,ds\right)^{\frac{1}{2}}\,.
	\end{align*}
Now, since by assumption $\frac{1}{w(s)}ds\leq C_g(a,b)\nu_{g}(s)ds$ on the interval $[a,b]$, we have
	\begin{align*}
		\|\text{II}\|_{2}&\leq\,W_{a,b}^{\frac{1}{2}}C_g(a,b)^{\frac{1}{2}}\left(\int_{0}^{\infty}\|\Delta_{\rho}^{\frac{1}{2}}\tilde{A}_{s,t}\|_{2}^{2}\nu_{g}(s)\,ds\right)^{\frac{1}{2}}\\
		&\overset{(2)}{\le }W_{a,b}^{\frac{1}{2}}C_g(a,b)^{\frac{1}{2}}\left(\gamma_{\rho}^{g}(A)-\gamma_{\Phi(\rho)}^{g}(\Phi(A))\right)^{\frac{1}{2}}.
	\end{align*}
The	inequality (2) above is justified from the rewriting of $\gamma_{\rho}^{g}(A)-\gamma_{\Phi(\rho)}^{g}(\Phi(A))$ as follows:
\begin{align*}
	&\gamma_{\rho}^{g}(A,A)-\gamma_{\Phi(\rho)}^{g}(\Phi(A),\Phi(A))\\
	&\qquad\qquad\qquad\qquad=\int_{0}^{\infty}\left(\langle A,\frac{R_{\rho}^{-1}}{s+\Delta_{\rho}}(A)\rangle\,-\langle \Phi(A), \frac{R_{\Phi(\rho)}^{-1}}{s+\Delta_{\Phi(\rho)}} (\Phi(A))\rangle \right)\nu_{g}(s)ds\\
	&\qquad\qquad\qquad\qquad=\int_{0}^{\infty}\langle A\rho^{-\frac{1}{2}+it},\,(s+\Delta_{\rho})^{-1}(A\rho^{-\frac{1}{2}+it}) \rangle \nu_{g}(s)\, ds\\
	&\qquad\qquad\qquad\qquad\quad -\int_{0}^{\infty}\,\langle A\rho^{-\frac{1}{2}+it},\,V_{\rho,t}(s+\Delta_{\Phi(\rho)})^{-1}V_{\rho,t}^{*}(A\rho^{-\frac{1}{2}+it}) \rangle \nu_{g}(s)\, ds.
\end{align*}
Then by Lemma \ref{lemma:contraction},
\begin{align*}
	\gamma_{\rho}^{g}(A,A)-\gamma_{\Phi(\rho)}^{g}(\Phi(A),\Phi(A))&\geq \int_{0}^{\infty}\,\langle \tilde{A}_{s,t},\,(s+\Delta_{\rho})(\tilde{A}_{s,t}) \rangle\nu_{g}(s)\,ds\\
	&\geq \int_{0}^{\infty}\,\langle \tilde{A}_{s,t},\,\Delta_{\rho}(\tilde{A}_{s,t}) \rangle\nu_{g}(s)\,ds\\
&\geq \int_{a}^{b}\|\Delta_{\rho}^{\frac{1}{2}}\tilde{A}_{s,t}\|_{2}^{2} \nu_{g}(s)\,ds
\end{align*}
from which (2) follows. This finishes the proof by adding up the bounds found for I, II and III.
\end{proof}

\begin{rem}{\rm The quantity
$h_1=
\tr(A^*\rho^{-1}A)^{1/2}= \gamma_\rho^{\RLD}(\rho)$ is the RLD metric, which satisfies the monotonicity. On the other hand,
\[\tr(\rho^{-2}A^*\rho A)=\lan A, \Delta_\rho R_\rho^{-1} A\ran  \]
corresponds to $\gamma^g$ with $g(x)=x$, which is not a monotone metric. So we have to also include  the term $\tr(\Phi(\rho)^{-2}\Phi(A)^*\Phi(\rho) \Phi(A))$ in the estimate .}
\end{rem}

We are now ready to prove the recovery of monotone metric.

\begin{theorem}\label{thm:monotone}
Let $\rho\in \cD_+(\cH)$ and $\Phi:\cB(\cH)\to \cB(\cK)$ be a quantum channel. Then for any $A\in \cB(\cH)$, the following are equivalent
\begin{enumerate}
\item[i)] $\gamma_\rho^g(A)=\gamma_{\Phi(\rho)}^g(\Phi(A))$ for all monotone metric $\gamma_g$.
\item[ii)] $\gamma_\rho^g(A)=\gamma_{\Phi(\rho)}^g(\Phi(A))$ for some regular monotone metric $\gamma_g$.
\item[iii)] $A=\cR_{\rho,\Phi}^t\circ \Phi(A)$ for some/all $t\in \mathbb{R}$ where $\cR_{\rho,\Phi}^t$ is the rotated Petz map.
\item[iv)] There exists a quantum channel $\cR$ such that $\rho=\cR\circ\Phi(\rho)$ and $A=\cR\circ \Phi(A)$.
\end{enumerate}
\end{theorem}
\begin{proof}The i) $\Rightarrow$ ii) and iii) $\Rightarrow$ iv) are trivial. The direction iv)  $\Rightarrow$ i) follows from monotonicity. The direction ii)  $\Rightarrow$ iii) follows from Lemma \ref{1abg}. Indeed, note that for a regular $g$, one can choose $w(s)=s$, and $W_{a,b},C_g(a,b)$ are finite for any $0<a<b<\infty$. Choosing $a=\frac{1}{b}$ arbitrarily small yields the equality iii).
\end{proof}

Recall that we say a quantum channel $\Phi$ is sufficient for a pair $\{\rho,\sigma\}$ of states whenever there exists a quantum channel $\cR$ such that $\cR\circ \Phi (\rho)=\rho$ and $\cR\circ \Phi (\sigma)=\sigma$.
\begin{cor}\label{cor:chi2}
Let $\rho,\sigma\in\cD(\cH)$ be two quantum states with $\operatorname{supp}(\sigma)\subseteq \operatorname{supp}(\rho)$.
For a quantum channel $\Phi:\cB(\cH)\to \cB(\cK)$, the following are equivalent
\begin{enumerate}
\item[i)] $\chi_g^2(\rho,\sigma)=\chi_g^2(\Phi(\rho),\Phi(\sigma))$ for all quantum $\chi^2$-divergence $\chi^2_g$;
\item[ii)] $\chi_g^2(\rho,\sigma)=\chi_g^2(\Phi(\rho),\Phi(\sigma))$ for some regular quantum $\chi^2$-divergence $\chi^2_g$;
\item[iii)] $\rho=\cR_{\sigma,\Phi}^t\circ \Phi(\rho)$ for some/all $t\in \mathbb{R}$ where $\cR_{\sigma,\Phi}^t$ is the rotated Petz map.
\item[iv)] $\Phi$ is sufficient for $\{\rho,\sigma\}$.
\end{enumerate}
\end{cor}
\begin{proof} This follows from Theorem \ref{thm:monotone} by choosing $A=\rho-\sigma$. Note that for non-faithful $\sigma$, we can always restrict the discussion on the support of $\sigma$.
\end{proof}

\begin{rem}{\rm 
The above recovery results does not hold for SLD and RLD metric
\[\gamma_\rho^{\RLD}(A)=\tr(A^*A\rho^{-1})\pl,\pl \gamma_\rho^{\SLD}(A)=\lan A, \bJ_\rho^{\SLD}(A) \ran \]
where $L=\bJ_\rho^{\SLD}(A)$ is the operator $L$ such that
\begin{align} A=\frac{1}{2}(L\rho +\rho L)\pl. \label{eq:L}\end{align}
It was argued in Section \ref{Sec:counterexample} that if $[A,L]\neq 0$,
\[ \gamma_{\cL(\rho)}^{\RLD}(\cL(A))=\gamma_{\cL(\rho)}^{\SLD}(\cL(A))=\gamma_{\rho}^{\SLD}(A)<\gamma_{\rho}^{\RLD}(A)\pl,\]
where $\cL$ is the pinching map for the spectrum of $L$. Thus $\gamma^{\SLD}$ is preserved under $\cL$ but $\cL$ is not sufficient for $\rho$ and $A$. For RLD metric, we note that 
\[\chi_{\RLD}^2(\rho,\sigma)=\gamma^{\RLD}_\sigma(\rho-\sigma)=\tr( (\rho-\sigma)^2\sigma^{-1})=\tr( \rho^2\sigma^{-1})-1\]
which is essentially the Petz-R\'enyi 2-divergence $Q_2(\rho\|\sigma)=\tr( \rho^2\sigma^{-1})$. It has been observed in \cite[Example 4.8]{hiai2017different} there exists states $\rho,\sigma$ and quantum channel $\Phi$ such that
\[ Q_2(\rho\|\sigma)=Q_2(\Phi(\rho)\|\Phi(\sigma))\pl,\]
but $\Phi$ is not sufficient for $\{\rho,\sigma\}$.
}
\end{rem}

\subsection{Approximate Recovery of monotone metrics}
In this section, we discuss approximate recovery bounds for specific metrics. Our starting point is a simple argument for general monotone metrics. Given a quantum channel $\Phi$, the data processing inequality
\begin{align}
    \label{eq:DPI} \lan \Phi(A), \bJ^g_{\Phi(\rho)}( \Phi(A)) \ran=\gamma_{\Phi(\rho)}^g(\Phi (A),\Phi (A)) \le \gamma_{\rho}^g(A,A)=\lan A, \bJ^g_\rho(A) \ran \pl.
\end{align}
can be interpreted as
\[ \Phi^\dagger \bJ_{\Phi(\rho)}^g \Phi\le \bJ_\rho^g\pl,\]
as positive operators on the Hilbert-Schmidt space $\mathcal{S}_2(\cH)$. By the operator anti-monotonicity of $x\mapsto x^{-1}$, we have
\[ (\Phi^\dagger \bJ_{\Phi(\rho)}^g \Phi)^{-1}\ge (\bJ_\rho^g)^{-1}\pl. \]
This leads to the following lemma.
 \begin{lemma}\label{lemma:simple}
For any density operator $\rho\in \cD_+(\cH)$ and $A\in \mathcal{B}(\mathcal{H})$,
\begin{align}
\gamma_\rho^g(A)-\gamma_{\Phi(\rho)}^g(\Phi(A))\ge\gamma_{\rho}^g(A-(\bJ_\rho^g)^{-1} \Phi^\dagger \bJ_{\Phi(\rho)}^g\Phi(A))\ge  \norm{A-\bJ_\rho^g \Phi^\dagger \bJ_{\Phi(\rho)}^g\Phi(A)}{1}^2
    \end{align}
 \end{lemma}
 \begin{proof}We have
 \begin{align*}
  \gamma_\rho^g( A) -\gamma_{\Phi(\rho)}^g(\Phi(A)) &=\lan A,\bJ_\rho^g A \ran- \lan \Phi(A), \bJ_{\Phi(\rho)}^g \Phi(A) \ran
  \\  &=\lan A, \bJ_\rho^g( A-(\bJ_\rho^g)^{-1} \Phi^\dagger \bJ_{\Phi(\rho)}^g\Phi(A))\ran
  \\  &\overset{(1)}{\ge} \lan A-(\bJ_\rho^g)^{-1} \Phi^\dagger  \bJ_{\Phi(\rho)}^g\Phi(A) , \bJ_\rho^g( A-(\bJ_\rho^g)^{-1} \Phi^\dagger  \bJ_{\Phi(\rho)}^g\Phi(A) )\ran
  \\&= \gamma_\rho^g\Big(A-(\bJ_\rho^g)^{-1} \Phi^\dagger  \bJ_{\Phi(\rho)}^g\Phi(A)\Big)
  \\ &\overset{(2)}{\ge} \norm{ A-(\bJ_\rho^g)^{-1} \Phi^\dagger  \bJ_{\Phi(\rho)}^g\Phi(A)}{1}^2
 \end{align*}
 Here the first inequality follows from
\begin{align*} \lan & A-(\bJ_\rho^g)^{-1} \Phi^\dagger  \bJ_{\Phi(\rho)}^g\Phi(A) , \bJ_\rho^g( A-(\bJ_\rho^g)^{-1} \Phi^\dagger  \bJ_{\Phi(\rho)}^g\Phi(A) )\ran
\\ &\quad =\lan A, \bJ_\rho^g( A-(\bJ_\rho^g)^{-1} \Phi^\dagger  \bJ_{\Phi(\rho)}^g\Phi )\ran-\lan (\bJ_\rho^g)^{-1} \Phi^\dagger  \bJ_{\Phi(\rho)}^g\Phi(A) , \bJ_\rho^g( A-(\bJ_\rho^g)^{-1} \Phi^\dagger  \bJ_{\Phi(\rho)}^g\Phi(A) )\ran
\\ &\quad =\lan A, \bJ_\rho^g( A-(\bJ_\rho^g)^{-1} \Phi^\dagger  \bJ_{\Phi(\rho)}^g\Phi )\ran -\lan  \Phi^\dagger  \bJ_{\Phi(\rho)}^g\Phi(A) ,  A\ran+ \lan \Phi^\dagger  \bJ_{\Phi(\rho)}^g\Phi(A),
(\bJ_\rho^g)^{-1} \Phi^\dagger  \bJ_{\Phi(\rho)}^g\Phi(A) )\ran
\end{align*}
and for the last two terms we have
\begin{align*}\lan \Phi^\dagger  \bJ_{\Phi(\rho)}^g\Phi(A),
(\bJ_\rho^g)^{-1} \Phi^\dagger  \bJ_{\Phi(\rho)}^g\Phi(A) \ran
&\le     \lan \Phi^\dagger  \bJ_{\Phi(\rho)}^g\Phi(A), \Big(\Phi^\dagger \bJ_{\Phi(\rho)}^g \Phi\Big)^{-1}  \Phi^\dagger  \bJ_{\Phi(\rho)}^g\Phi(A) \ran
\\ &=  \lan A, \Phi^\dagger  \bJ_{\Phi(\rho)}^g\Phi(A) \ran\,.
\end{align*}
The second inequality was found in \cite{temme2010chi}, the proof of which we include for completeness: it is sufficient to consider the Bures metric
\[\gamma_\rho^{\text{Bures}}(X)=\lan X,\bJ^b_\rho (X) \ran \pl, \pl \bJ^b_\rho=2(L_\rho+R_\rho)^{-1} \pl,\]
since $\gamma_\rho^{\text{Bures}}(X)\le \gamma_\rho^{g}(X)$ for any $g$. Then
\begin{align*} \norm{X}{1}^2=&\sup_{U}|\tr(XU)|^2=\sup_{U} |\lan U,X \ran|^2
\\=& \sup_{U}\lan (\bJ_\rho^b)^{-\frac{1}{2}}U, (\bJ_\rho^b)^{\frac{1}{2}}X \ran
\\ \le& |\sup_{U} \lan U, (\bJ_\rho^b)^{-1}(U)\ran| \cdot  \lan X, \bJ_\rho^b(X)\ran
\\= & |\sup_{U} \frac{1}{2}\tr( U^* U\rho+U^*\rho U)\ran|  \cdot  \gamma_\rho^{\text{Bures}}(X)=\gamma_\rho^{\text{Bures}}(X).
\end{align*}
That completes the proof.
 \end{proof}

 The above estimate is nice but not necessarily gives a recovery bound. The reason is that the map
 \[ (\bJ_\rho^g)^{-1} \Phi^\dagger  \bJ_{\Phi(\rho)}^g\]
 is not necessarily a channel. Indeed, such a map is always trace preserving because the adjoint is unital
 \[ \bJ_{\Phi(\rho)}^g \Phi (\bJ_\rho^g)^{-1} (1)=\bJ_{\Phi(\rho)}^g (\Phi(\rho))=1\pl. \]
 But the map $(\bJ_\rho^g)^{-1}$ may not be positive, as it is the case for the Bures metric
 \[ (\bJ_\rho^b)^{-1}(A)=\frac{1}{2}(\rho A+A\rho)\]
 Nevertheless, the situation simplifies in the case of the $\frac{1}{2}$-metric
\[ \gamma^{s}_\rho(A,B)=\tr(A^*\rho^{-\frac{1}{2}}B \rho^{-\frac{1}{2}})= \lan A, \bJ_{\rho}^{s}B \ran \pl, \]
where the multiplication operator is
\[\bJ_{\rho}^{s}(A)=\rho^{-\frac{1}{2}}A\rho^{-\frac{1}{2}}\pl.\]
The inverse operator is
\[ (\bJ_{\rho}^{s})^{-1}(A)=\rho^{\frac{1}{2}}A\rho^{\frac{1}{2}}\pl.\]
 Then it is clear that for $\frac{1}{2}$-metric
 \[ (\bJ_\rho^s)^{-1} \Phi^\dagger  \bJ_{\Phi(\rho)}^s=\mathcal{R}_{\rho,\Phi}\]
 gives the Petz recovery map. Thus we have the following simple recovery bound for  $\frac{1}{2}$-metric
 \begin{theorem}\label{thm:1/2}
For any density operator $\rho\in \cD_+(\cH)$ and $A\in \mathcal{B}(\mathcal{H})$,
\begin{align}
\gamma_\rho^s(A)-\gamma_{\Phi(\rho)}^s(\Phi(A))\ge\gamma_{\rho}^s(A-\mathcal{R}_{\rho,\Phi}\circ\Phi(A))\ge  \norm{A-\mathcal{R}_{\rho,\Phi}\circ\Phi(A)}{1}^2
    \end{align}
    In terms of $\chi^2_{1/2}$ divergence, for two quantum states $\rho$ and $\sigma$ with $\operatorname{supp}(\sigma)\le\operatorname{supp}(\rho)$,
    \[\chi^2_{1/2}(\sigma,\rho)-\chi^2_{1/2}(\Phi(\sigma),\Phi(\rho))\ge \norm{\sigma-\mathcal{R}_{\rho,\Phi}\circ\Phi(\sigma)}{1}^2\]
 \end{theorem}
\noindent A weaker inequality \[\norm{\sigma-\cR_{\rho,\Phi}\circ \Phi(\sigma)}{1}^2\le \norm{\sigma^{-1}}{\infty}\Big(\chi^2_{\frac12}(\sigma,\rho)-\chi^2_{\frac12}(\Phi(\sigma),\Phi(\rho))\Big)\]
 was obtained in \cite{cree2022approximate}. Here we removed the singularity term $ \norm{\sigma^{-1}}{\infty}$.

 We now discuss some approximate recovery results using Lemma \ref{1abg}. Recall the Bogolyubov-Kubo-Mori metric is
 \[ \gamma_{\rho}^{\BKM}(A)=\int_{0}^\infty\tr(A^*(\rho+s)^{-1} A(\rho+s)^{-1})ds \,.\]
The associated measure is $\nu(s)ds=\frac{1}{1+s}ds$.
\begin{corollary}\label{BKMmain}
 For all $\rho\in\mathcal{D}_+(\mathcal{H})$ and $A\in\mathcal{B}(\mathcal{H})$, we have
	\begin{align*}
		\gamma_{\rho}^{\BKM}(A)-\gamma_{\Phi(\rho)}^{\BKM}(\Phi(A))\geq \sup_{0<\epsilon<\frac{1}{2}}\left(\frac{\pi}{\cosh(\pi t)}\frac{\|A-\mathcal{R}_{\rho,\Phi}^{t}(\Phi(A))\|_{1}}{K(\rho,A,\epsilon)}\right)^{\frac{4}{1-2\epsilon}},
	\end{align*}
	where  \begin{align*}
		K(\rho,A,\epsilon):=4\tr(A^{*}\rho^{-1}A)^{\frac{1}{2}}+2\tr(\rho^{-2}A^{*}\rho A)^{\frac{1}{2}}+2\tr(\Phi(\rho)^{-2}\Phi(A)^{*}\Phi(\rho)\Phi(A))^{\frac{1}{2}}+1+(\epsilon e)^{-\frac{1}{2}}\,.
	\end{align*}
\end{corollary}
\begin{proof}

By choosing $w(s)=1+s$, Lemma \ref{1abg} applies with the global constant $C_g=1$. With the notations of Lemma \ref{1abg}, we hence have that, by choosing $\delta:=\min\{h_3
	,\,1\}$, $a=b^{-1}=\delta$,
	 \begin{align*}
			\|A-\mathcal{R}_{\rho,\Phi}^{t}(\Phi(A))\|_{1}&\leq\frac{\cosh(\pi t)}{\pi} \delta^{\frac{1}{2}}\left (4h_{1}+4h_{2}+\Big(-2\ln\delta -\delta+\frac{1}{\delta}\Big)^{\frac{1}{2}}\delta^{\frac{1}{2}}\right)\\
			&\leq\frac{\cosh(\pi t)}{\pi} \delta^{\frac{1}{2}}\left (4h_{1}+4h_{2}+\Big(-2\ln\delta+\frac{1}{\delta}\Big)^{\frac{1}{2}}\delta^{\frac{1}{2}}\right)\\
&\overset{(1)}{\le} \frac{\cosh(\pi t)}{\pi}\delta^{\frac{1}{2}}\left( 4h_{1}+4h_{2}+(-2\ln \delta)^{\frac{1}{2}}\delta^{\frac{1}{2}}+1\right)
		\end{align*}
where in $(1)$ we have used the two-point inequality $(a+b)^{\frac{1}{2}}\le a^{\frac{1}{2}}+b^{\frac{1}{2}}$ for any $a,b\ge 0$. Finally, since $2\ln(\delta)+(\epsilon e)^{-1}\delta^{-2\epsilon}\ge 0$ for any $\epsilon>0$ and $\delta\in(0,1]$.
		Together with the fact that $\delta^{-\epsilon}\geq 1$ for $\delta\in(0,1]$ and $\epsilon>0$, we obtain that
		\begin{align*}
			\|A-\mathcal{R}_{\rho,\Phi}^{t}(\Phi(A))\|_{1}\leq \frac{\cosh(\pi t)}{\pi}\delta^{\frac{1}{2}-\epsilon}\left(4h_{1}+4h_{2}+(\epsilon e)^{-\frac{1}{2}} \right),
		\end{align*}
		which completes the proof.
\end{proof}

Another example is the $\al$-metric
 \[ \gamma_{\rho}^{(\al)}(A)=\tr(A^*\rho^{\al-1} A\rho^{-\al}) \]
The associated measure is $\nu(s)ds=\frac{\sin\pi \al}{\pi}s^\al ds$.

\begin{corollary}
 Let $\alpha\in(0,\frac{1}{2})$. Then, for all $\rho\in\mathcal{D}_+(\mathcal{H})$ and $A\in\mathcal{B}(\mathcal{H})$, we have
 \begin{align*}
\gamma_{\rho}^{(\al)}(A)-\gamma_{\Phi(\rho)}^{(\al)}(\Phi(A))\geq \left(\frac{\pi}{\cosh(\pi t)}\frac{\|A-\mathcal{R}_{\rho,\Phi}^{t}(\Phi(A))\|_{1}}{K(\rho,A,\epsilon)}\right)^{\frac{2}{\al}}\,,
\end{align*}
where
\begin{align*}K(\rho,A,\epsilon)=&4\tr(A^{*}\rho^{-1}A)^{\frac{1}{2}}+2\tr(\rho^{-2}A^{*}\rho A)^{\frac{1}{2}}+2\tr(\Phi(\rho)^{-2}\Phi(A)^{*}\Phi(\rho)\Phi(A))^{\frac{1}{2}}\\ &+\sqrt{\frac{\pi}{\al \sin(\pi \alpha)}}.\end{align*}
\end{corollary}
 \begin{proof}
 We take $w(s)=s^\al$ and $C_\al=\frac{\pi}{\sin(\pi \alpha)}$ so that $W_{a,b}=\int_{a}^bx^{\al-1}dx= \frac{1}{\al}(b^{\al}-a^{\al})$.
Then by Lemma \ref{1abg}, we have that, given $\delta:=\min\{h_3
	,\,1\}$,
\begin{align*}
&\|A-\mathcal{R}_{\rho,\Phi}^{t}(\Phi(A))\|_{1}\\
&\qquad \leq\frac{\cosh(\pi t)}{\pi} \Big(4h_{1}\delta^{1/2}+4h_{2}\delta^{1/2}+\sqrt{\al^{-1}C_\al(\delta^{-\al}-\delta^{\al})\delta}\Big)\\
&\qquad \leq\frac{\cosh(\pi t)}{\pi} \Big(4h_{1}+4h_{2}+\sqrt{\frac{\pi}{\al\sin(\pi \alpha)} }\Big)\delta^{\frac{\al}{2}}\,.\qedhere
\end{align*}
\end{proof}
For $\al=\frac{1}{2},t=0$, the above bound is of course worse than our Theorem \ref{thm:1/2}.

\section{Sufficiency of quantum Fisher information}

% Consider the monotonic metric
% \begin{align}
% 	\gamma^{g}_{\rho}(A,B)=\int_{0}^{\infty} \sigma_{g}(s)\langle A, \,\frac{R_{\rho}^{-1}}{s+\Delta_{\rho}}(A)\rangle\,ds,\label{lrrm}
% \end{align}
% where $\sigma_{g}(s):=N_{g}(s)+s^{-1}N_{g}(s^{-1})$ with  $$N_{g}(s)ds=(b_{g}+c_{g})\delta(s)ds+d\nu_{g}(s),$$

In this section, we discuss sufficiency of quantum Fisher information. Let $(a,b)$ be an interval and $\{\rho_\theta\}_{\theta\in (a,b)}$ be a smooth one-parameter family of quantum states. Given a monotone metric $\gamma^g$ in Definition \ref{def:gamma}, the associated quantum Fisher information of the family $\{\rho_\theta\}$ is defined as
\[ I_\rho^g(\theta)=\gamma_{\rho_\theta}^g(\dot{\rho}_\theta)\pl,\]
where $\dot{\rho}_\theta=\frac{d}{d\theta}\rho_\theta$ is the derivative of $\rho_\theta$ w.r.t~the parameter $\theta$.
It is inherited from $\gamma^g$ that $I^g$ satisfies the data processing inequality: for any quantum channel $\Phi$
\[ I^g_\rho(\theta)\ge I^g_{\Phi(\rho)}(\theta)\]
where the right-hand side is the Fisher information of the family $\theta\mapsto \Phi(\rho_\theta)$.
Recall that we say a channel $\Phi$ is sufficient for $\{\rho_\theta\}_{\theta\in (a,b)}$ if there exists a recovery channel $\cR$ such that $\cR\circ \Phi(\rho_\theta)=\rho_\theta$ for any $\theta\in (a,b)$.

\begin{theorem}\label{Thm5}
Let $(\rho_\theta)_{\theta\in (a,b)}$ be a smooth family of full-rank quantum states in $\mathcal{D}(\mathcal{H})$. A quantum channel $\Phi: \mathcal{B}(\mathcal{H})\to \mathcal{B}(\mathcal{\cK})$ is sufficient for $\{\rho_\theta\}_\theta$ if and only if
\[I^g_\rho(\theta)=I^g_{\Phi(\rho)}(\theta) \pl, \pl \forall \theta\pl \]
for all/some regular quantum Fisher information $I^g_{\rho}$. Moreover, the recovery map $\mathcal{R}$ can be chosen as the (rotated) Petz recovery map $\mathcal{R}_{\rho_o,\Phi}^t$ for any $o\in (a,b)$.
\end{theorem}
\begin{proof}  As Theorem \ref{thm:monotone} is valid on the support of $\rho_\theta$, $I^g_\rho(\theta)=I^g_{\Phi(\rho)}(\theta)$ implies that for any $t\in\mathbb{R}$ and $\theta\in (a,b)$
\[\dot{\rho}_\theta=\mathcal{R}_{\rho_\theta,\Phi}^{t}({\Phi(\dot{\rho}_\theta}))\,.\]
Thus for each $t\in\mathbb{R}$,
\[\rho_\theta^{-\frac{1}{2}+it}\dot{\rho}_\theta\rho_\theta^{-\frac{1}{2}-it}=\Phi^*\Big(\Phi(\rho_\theta)^{-\frac{1}{2}+it}{\Phi(\dot{\rho_\theta})}\Phi(\rho)_\theta^{-\frac{1}{2}-it}\Big)\]
Recall that
\[ \frac{d}{d\theta}\log\rho_\theta=\int_{\mathbb{R}}\rho_\theta^{-\frac{1+it}{2}}\dot{\rho}_\theta\rho_\theta^{-\frac{1-it}{2}}d\beta(t)\]
where $t$ is integrating over the probablity measure $d\beta(t)=\frac{\pi }{2(\cosh(\pi t)+1)}dt$, for example see
\cite{sutter2017multivariate}. Thus we have
\[ \frac{d}{d\theta}\log\rho_\theta=\Phi^*\Big(\frac{d}{d\theta}\log\Phi(\rho_\theta)\Big)\,.\]
Integrating the above equality over any finite interval $[o,\theta]\subset (a,b)$, we have
\[ \log\rho_\theta-\log\rho_o=\Phi^*( \log\Phi(\rho_\theta)-\log\Phi(\rho_o))\,.\]
We therefore obtain
\[D(\rho_\theta\|\rho_o)=D(\Phi(\rho_\theta)\|\Phi(\rho_o))\]
which implies $\rho_{\theta}=\mathcal{R}_{\rho_o,\Phi}^t\circ \Phi(\rho_\theta)$ for any $\theta$ and $t\in \mathbb{R}$.
\end{proof}

The above result can be easily extended to the case of multivariate parameters. Let $\Theta\subset \mathbb{R}^n$ be a simple connected region and
$\{\rho_\theta\}_{\theta\in \Theta}$ be a smooth family of quantum states. The Fisher information matrix is defined as
\[I^g_\rho(\theta):=[I^g_{\rho}(\theta)]_{ij}=\left[\gamma_{\rho_\theta}^g(\partial_i \rho_\theta, \partial_j \rho_\theta) \right]_{1\le i,j\le n}\pl.\]
$I^g_\rho(\theta)$ is a (real) $n\times n$ positive matrix. It is known that for any quantum channel $\Phi$
\[ I^g_\rho(\theta)\ge I^g_{\Phi(\rho)}(\theta)\]
as positive semi-definite matrix for every $\theta$.

\begin{theorem} Let $(\rho_\theta)_{\theta\in \Theta}$ be a smooth family of full-rank quantum states in $\mathcal{D}(\mathcal{H})$.
A quantum channel $\Phi$ is sufficient for $\{\rho_\theta\}$ if and only if
\[I^g_\rho(\theta)=I^g_{\Phi(\rho)}(\theta) \pl ,\pl \forall \theta\pl,\]
for all/some regular quantum Fisher information $I^g$.
\end{theorem}
\begin{proof}Take a smooth path $\eta:[0,1]\to \Theta$ such that $\eta(0)=o$ and $\eta(1)=\theta$.  Then $\rho_s:=\rho_{\eta(s)}$ is a one parameter family of states and
\[ \dot{\eta}(s)=D_{\dot{\eta}(s)}\rho_\theta= \sum_{j}\dot{\eta}_j(s)\partial_j \rho_\theta\]
where $\dot{\eta}(s)$ is the derivative of $\eta$ at $s$. Then
\[ I^g_\rho(s)=\gamma_{\rho_s}(D_{\dot{\eta}(s)}\rho_\theta,D_{\dot{\eta}(s)}\rho_\theta)\pl.\]
If \[I_\rho(\theta)=I_{\Phi(\rho)}(\theta) \]
as positive semi-definite matrix, we have
\[I_\rho(s)=  \sum_{ij}[I_{\rho}(\theta)]_{ij}\dot{\eta}_i(s)\dot{\eta}_j(s)
=\sum_{ij}[I_{\Phi(\rho)}(\theta)]_{ij}\dot{\eta}_i(s)\dot{\eta}_j(s)=I_{\Phi(\rho)}(s)
\]
where the right hand side is the QFI for the family $\Phi(\rho)_s:=\Phi(\rho)_{\eta(s)}$. Then the assertion follows from the one dimensional case.
\end{proof}

We now discuss a recovery bound using BKM Fisher information. Recall the BKM metric is
\[\gamma_\rho^{\BKM}(X)=\lan X,\bJ_\rho^{\BKM}(X)\ran = \int_{0}^\infty\tr(X^*(\rho+s)^{-1}X(\rho+s)^{-1})ds \pl. \]
The inverse map is
\[ (\bJ_\rho^{\BKM})^{-1}(X)=\int_{0}^1 \rho^{t}X\rho^{1-t}dt\]
It is easy to see that if $\rho\le C\sigma$ for $C>0$, then
\[ \lan X,  (\bJ_\rho^{\BKM})^{-1} X\ran \le  C\lan X,  (\bJ_\sigma^{\BKM})^{-1} X\ran \]
Note that here the optimal constant $C$ is the $D_{\max}$ relative entropy up to a logarithm
\begin{align*}D_{\max}(\rho\|\sigma)=\log \inf \{C>0 \pl |  \rho\le C\sigma\} \end{align*}

 \begin{theorem}
 Let $\{\rho_\theta\}_{\theta\in [0,1]}$ be a smooth family of quantum states.
 Then for any quantum channel $\Phi$, denoting $\Phi(\rho)_\theta:=\Phi(\rho_\theta)$,
\begin{align}\label{eq:approximate} D(\rho_1\|\rho_0)-D(\Phi(\rho)_1\|\Phi(\rho)_0)\le \int_{0}^1 e^{\frac{1}{2}D_{\max}(\rho_1\|\rho_\theta)}\sqrt{ I_\rho^{\BKM}(\theta)-I_{\Phi(\rho)}^{\BKM}(\theta) } \,d\theta\pl. \end{align}
 In particular, if $\rho_\theta\ge \la 1$ for any $\theta$,
 \[ D(\rho_1\|\rho_0)-D(\Phi(\rho)_1\|\Phi(\rho)_0)\le \lambda^{-\frac12}\int_{0}^1 \sqrt{ I_\rho^{\BKM}(\theta)-I_{\Phi(\rho)}^{\BKM}(\theta) }\, d\theta\pl.\]
 \end{theorem}
 \begin{proof} In the following, we use the short notations $\gamma=\gamma^{\BKM}$, $I=I^{\BKM}$ and $\bJ=\bJ^{\BKM}$.
By Lemma \ref{lemma:simple}, we have
\begin{align*} I_\rho(\theta)-I_{\Phi(\rho)}(\theta)\ge&\norm{ \bJ_{\rho_\theta}^{-\frac{1}{2}} (\bJ_{\rho_\theta}(\dot{\rho}_\theta)- \Phi^\dagger  \bJ_{\Phi({\rho_\theta})}\Phi(\dot{\rho}_\theta)) }{2}^{{2}}
\\ \ge& e^{-D_{\max}({\rho_1\|\rho_\theta})}\norm{ \bJ_{\rho_1}^{-\frac{1}{2}} (\bJ_{\rho_\theta}(\dot{\rho}_\theta)- \Phi^\dagger  \bJ_{\Phi({\rho_\theta})}(\dot{\Phi(\rho)_\theta})) }{2}^{{2}}
\\ =& e^{-D_{\max}({\rho_1\|\rho_\theta})}\norm{ \bJ_{\rho_1}^{-\frac{1}{2}} ( \dot{(\log\rho_\theta)}- \Phi^\dagger  (\dot{\log\Phi(\rho)_\theta}) }{2}^{{2}}
\end{align*}
Then
\begin{align*} &\left\|\bJ_{\rho_1}^{-\frac{1}{2}}\Big( \log \rho_1-\log \rho_0- \Phi^\dagger( \log \Phi(\rho_1)- \log \Phi(\rho_0))\Big)\right\|_2 \\ &\qquad\qquad \le \int_0^1 \norm{{ \bJ_{\rho_1}^{-\frac{1}{2}}(\dot{(\log\rho_\theta)}-\Phi^\dagger \dot{(\log\Phi(\rho)_\theta)})}}{2}d\theta\\ &\qquad\qquad \le \int_0^1  e^{\frac{1}{2}D_{\max}({\rho_1\|\rho_\theta})}\sqrt{I_\rho(\theta)-I_{\Phi(\rho)}(\theta)} \,d\theta\,.
\end{align*}
Therefore,
\begin{align*}
D(\rho_1\|\rho_0)-D(\Phi(\rho)_1\|\Phi(\rho)_0)= &\tr\Big(\rho_1 \big (\log \rho_1-\log \rho_0- \Phi^\dagger( \log \Phi(\rho_1)- \log \Phi(\rho_0)\big)\Big)\\ \le &
\left\| \bJ_{\rho_1}^{-\frac{1}{2}}\Big(\log \rho_1-\log \rho_0- \Phi^\dagger\big( \log \Phi(\rho_1)- \log \Phi(\rho_0)\big)\Big) \right\|_{2}
\\ \le &\int_0^1  e^{\frac{1}{2}D_{\max}({\rho_1\|\rho_\theta})}\sqrt{I_\rho(\theta)-I_{\Phi(\rho)}(\theta)} \,d\theta \qedhere
\end{align*}
 \end{proof}
 \begin{remark}{The above estimate gives a recovery bound via
 \begin{align}\label{eq:approximate}
D(\rho\|\si)- D(\Phi(\rho)\|\Phi(\si))\ge \frac{1}{4} \norm{\rho-\mathcal{R}_{\sigma,\Phi}^{\operatorname{uni}}\circ \Phi(\rho)}{1}^2,
\end{align}
where $\mathcal{R}_{\sigma,\Phi}^{\operatorname{uni}}$ is the universal recovery map. In the case when $\rho_1=\rho_t$ is a small perturbation of $\rho_0$ and $t\to 0$, the estimate \eqref{eq:approximate} is not of optimal asymptotic order,  because
\[ D(\rho_t\|\rho_0)-D(\Phi(\rho)_t\|\Phi(\rho)_0)=(I_\rho(0)-I_{\Phi(\rho)}(0))t^2+O(t^3) \]}
 \end{remark}

\section{Recoverability of asymmetry}
In general, the resource theory of asymmetry can be defined for all symmetry groups. Let $G$ be a compact Lie group and $\mathfrak{g}$ be its Lie algebra. We assume each system under consideration has a given unitary representation of $G$.  To quantify the asymmetry of a state $\rho$ with respect to symmetry $G$, we consider the QFI
metric for the family of states
$\rho_g=U(g) \rho U^*(g)$,
 where $U$ is the unitary representation on the system. To express the metric, it suffices to consider a neighborhood of the identity element $e$ of the Lie group, which can be smoothly parameterized by $n$ real parameters,
  denoted by $\Theta=(\theta_1,\cdots, \theta_n)\in \Omega \subset \mathbb{R}^n$, where $n=\dim(\mathfrak{g})$. Then for the parametrized family $\rho_\Theta=U(g(\Theta)) \rho U^*(g(\Theta))$, the QFI matrix at $\rho$ is the $n\times n$ matrix
\begin{equation}\label{IM:matrix}
I_{G}(\rho):=\Big[\gamma_\rho(\partial_i \rho_g , \partial_j \rho_g)\Big|_{g=e}\Big]_{1\le i,j\le n}= \Big[\gamma_{\rho}([L_i, \rho], [L_j, \rho])\Big]_{1\le i,j\le n}\ ,
\end{equation}
   where  $\partial_i \rho_g=\frac{\partial \rho_{g(\Theta)}}{\partial \theta_i} $ is the partial derivatives and  the skew-Hermitian operators
 \begin{equation}
 L_i=\frac{\partial U(g(\Theta))}{\partial \theta_i}\Big|_{g=e} \ \ \ \  : i=1,\cdots, n\ ,
\end{equation}
 form a basis for the representation of the Lie algebra $\mathfrak{g}$ induced by the unitary representation of $G$.

 As an example, one can choose the local parametrization by the exponential map such that for $\Theta=(\theta_1,\cdots, \theta_n)$, the corresponding unitary is $U(g(\Theta))=\exp({\sum_i\theta_i L_i})$. \footnote{Note that in the case of compact connected Lie groups, the exponential map from the Lie algebra to the Lie group is surjective.}.  For instance, in the case of $SO(3)$ symmetry corresponding to rotations in $3D$ space, the parametrization can be chosen such that operators $L_1, L_2, L_3$ become the angular momentum operators in $x , y, z$ directions. Then, the unitary $\exp({\phi \sum_i n_i L_i})$ for a  real vector $\hat{n}=(n_1,n_2,n_3)$ with the normalization  $\|\hat{n}\|_2=1$, corresponds to the rotation by angle $\phi$ around the axis $\hat{n}$. In this case the QFI metric will be a $3\times 3$ real matrix that determines the sensitivity of state $\rho$ under rotations around $x, y, z$ axes.
 
 As we saw in Eq.(\ref{IM:def1}) in the case of time translation symmetry,  the value of QFI with respect to the time parameter $t$ is constant for the entire family of time-evolved versions of state. However, this is not the case for a general group $G$, where QFI matrix will depend on the group element $g\in G$. Nevertheless, it turns out that for compact connected Lie groups, the QFI matrix $I_G(\rho)$ determines the QFI matrix for the entire family $\rho_g=U(g)\rho U^*(g): g\in G$.
Let $\eta=(\eta_1,\cdots, \eta_n)\in\mathbb{R}^n$ be a local coordinate at point $g\in G$.
Then, the QFI matrix defined relative to this coordinate system   is related to the QFI matrix at the identity via the congruence transformation
\begin{equation}\label{trans4}
\Big[\gamma_\rho(\partial_i \rho_g , \partial_j \rho_g)\Big]_{i,j}= V^T(g)\Big[\gamma_\rho(\partial_i \rho_g , \partial_j \rho_g)\big|_{g=e}\Big]_{i,j} V(g)=V^T(g)I_{G}(\rho)V(g)\ ,
\end{equation}
where  $\partial_i \rho_g=\partial \rho_{g(\Phi)}/\partial \eta_i $
and  $V(g)$ is an invertible $n\times n$ real matrix defined by equation
\begin{equation}
U^*(g)\frac{\partial }{\partial \eta_i}U(g(\Phi))=  \sum_{r=1}^n V_{ri}(g) L_r\ .
\end{equation}
Note that here the matrix $V$ is only determined by the local coordinate (or equivalently the basis in Lie algebra) but independent of the representation $U$ \footnote{The exact form of matrix $V(g)$ can be obtained, e.g., via equation $$\frac{d}{ds} \exp({X(s)})=\exp({X(s)}) \frac{1-\exp({-\text{ad}_{X(s)}})}{\text{ad}_{X(s)}}\frac{d}{ds} X(s)\ ,$$ where  $\text{ad}_{X} (Y)=[X, Y]$.}.

A CPTP map $\cE:\cB(\cH_A)\to \cB(\cH_B)$ respects this symmetry, or is called covariant, if it satisfies the covariance condition
 \begin{equation}\label{cov2}
  \mathcal{E}\big(U_A(g) (\cdot) U^*_A(g)   \big)=U_B(g)\mathcal{E}( \cdot) U^*_B(g)  \ , \forall g\in G
\end{equation}
where $U_A$ and $U_B$ are the given representations of $G$ on the input system $\cH_A$ and output systems $\cH_B$, respectively.
Then, for the family $\rho_g=U_A(g)\rho U_A^*(g), \rho \in \cB(\cH_A)$,
\[ \cE(\rho_g)=\cE\big(U_A(g)\rho U_A^*(g)\big) = U(g)\cE(\rho) U^*(g) =\cE(\rho)_g  \]
The monotonicity of the QFI matrix under data processing then implies 
\begin{equation}
I_G(\rho)\ge I_G(\mathcal{E}(\rho))  \ .
\end{equation}
as matrices, for any covariant map $\mathcal{E}$ and any QFI metric $I_g$. In the case of regular QFI metrics, our result in Theorem \ref{Thm5} implies that conservation of QFI metric guarantees reversibility.
 \begin{theorem}\label{IM:Thm2}
Let $G$ be a compact connected Lie group and let $\cH_A$ and $\cH_B$ be two systems with representations $U_A$ and $U_B$ of the group $G$. Let $\rho\in \cB(\cH_A)$ be a full rank density operator and $\cE:\cB(\cH_A)\to \cB(\cH_B)$ be a covariant quantum channel. 
Then, this process is reversible with a covariant operation $\mathcal{R}$ from $B$ to $A$, such that $\mathcal{R}(\cE (\rho))=\rho$,  if and only if
\[I_{G}(\rho)=I_{G}(\cE(\rho))\]
where  $I_G$ can be the QFI matrix in Eq.(\ref{IM:matrix}) defined by some/all regular QFI metric
 \end{theorem}
%Note that, the representation $U$ and $U'$, in general, can be different. The matrix elements  $[I_G(\rho')]_{ij}$ of the output QFI matrix are given by  $\gamma_{\rho'}([L'_i, \rho'], [L'_j, \rho'])\Big]$, where  $L'_i=\frac{\partial U'(g(\Theta))}{\partial \theta_i}|_{g=e}$.

%To simplify the notation in the following we assume the representation of the group on the input and output systems  are identical and is denoted by $U$.

The proof of this theorem is similar to the case of time-translation symmetry for the special case of a periodic representation, in which it suffices to consider the average Petz map Eq. \ref{eq:Ravg} with the time $T$ being the period. 

\begin{proof}By the congruence transformation \eqref{trans4}, we see $I_{G}(\rho)=I_{G}(\cE(\rho))$ implies the QFI matrix
\[ I_\rho(g)=I_{\cE(\rho)}(g) \pl,\]
are preserved at $\rho_g=U_A(g)\rho  {U_A(g)}^*$ for every $g\in G$. By Theorem \eqref{Thm5}, this implies the Petz recovery map $\mathcal{R}_{\rho,\mathcal{E}}$ satisfies
\begin{align}
\mathcal{R}_{\rho,\mathcal{E}}\big(U_B(g)\cE(\rho) {U_B(g)}^*\big)=U_A(g)\rho\  {U_A(g)}^*\ .
\end{align}
In general, this Petz  map $\mathcal{R}_{\rho,\mathcal{E}}$ is not covariant. By twirling this map with the uniform (Haar) measure over the group $G$, we obtain the map
\begin{align}
\mathcal{R}_{\text{avg}}(\cdot) :=\int_G \ \ {U_A(g)}^* \mathcal{R}_{\rho_B,\mathcal{E}}\left(U_B(g)(\cdot) {U_B(g)}^*\right)U_A(g) d\mu(g) \ ,
  \end{align}
which is covariant and satisfies $\mathcal{R}_{\text{avg}}(\cE(\rho))=\rho$. This completes the proof.  
\end{proof}

\bibliographystyle{alpha}
\bibliography{tube,QISE}
\end{document}